\documentclass[preprint,number]{elsarticle}

\usepackage{amsthm,amsmath,amssymb}
\usepackage{booktabs}
\usepackage{paralist}
\usepackage{wrapfig}
\usepackage{placeins}
\usepackage[ruled,lined,linesnumbered]{algorithm2e}
\usepackage{url}
\usepackage[autolanguage,np]{numprint}
\usepackage{subcaption}
\usepackage{xcolor}
\usepackage{soul}
\usepackage{colortbl}
\usepackage{mwe}
\usepackage{xcolor}
\usepackage{ifthen}

\newboolean{cameraready}
\setboolean{cameraready}{true} % Change to true for camera-ready

\ifthenelse{\boolean{cameraready}}
  {\colorlet{maincolor}{black}} % Camera-ready color
  {\colorlet{maincolor}{blue}}  % Draft color

\usepackage{natbib}
\newtheoremstyle{thmlemcorr}{10pt}{10pt}{\itshape}{}{\bfseries}{.}{10pt}{{\thmname{#1}\thmnumber{ #2}\thmnote{ (#3)}}}
\newtheoremstyle{thmlemcorr*}{10pt}{10pt}{\itshape}{}{\bfseries}{.}\newline{{\thmname{#1}\thmnumber{ #2}\thmnote{ (#3)}}}
\newtheoremstyle{remexample}{10pt}{10pt}{}{}{\bfseries}{.}{10pt}{{\thmname{#1}\thmnumber{ #2}\thmnote{ (#3)}}}
\newtheoremstyle{ass}{10pt}{10pt}{}{}{\bfseries}{.}{10pt}{{\thmname{#1}\thmnumber{ A#2}\thmnote{ (#3)}}}

\theoremstyle{thmlemcorr}
\newtheorem{theorem}{Theorem}
\newtheorem{definition}[theorem]{Definition}

\theoremstyle{thmlemcorr*}
\newtheorem{theorem*}{Theorem}
\newtheorem{lemma*}[theorem]{Lemma}
\newtheorem{corollary*}[theorem]{Corollary}
\newtheorem{proposition*}[theorem]{Proposition}
\newtheorem{problem*}[theorem]{Problem}
\newtheorem{conjecture*}[theorem]{Conjecture}
\newtheorem{definition*}[theorem]{Definition}
\newtheorem{assumption*}[theorem]{Assumption}

\theoremstyle{remexample}
\newtheorem{remark}[theorem]{Remark}

\theoremstyle{ass}

\newcommand{\R}{\mathbb{R}}

\definecolor{forestgreen}{HTML}{228B22}

\renewcommand{\epsilon}{\varepsilon}
\renewcommand{\phi}{\varphi}
\newcommand{\methodname}{FAIM\xspace}
\newcommand{\compas}{COMPAS\xspace}

\SetAlFnt{\footnotesize}
\SetKwInOut{AlgInput}{input}
\SetKwInOut{AlgOutput}{output}
\SetKwComment{AlgComment}{// }{}

\SetCommentSty{mycommfont}
\SetAlCapFnt{\footnotesize}
\SetAlCapNameFnt{\footnotesize}
\newcommand{\nosemic}{\renewcommand{\@endalgocfline}{\relax}}% Drop semi-colon ;
\newcommand{\dosemic}{\renewcommand{\@endalgocfline}{\algocf@endline}}% Reinstate semi-colon ;
\let\oldnl\nl% Store \nl in \oldnl
\newcommand{\nonl}{\renewcommand{\nl}{\let\nl\oldnl}}% Remove line number for one line

\newcommand{\ie}{i.e., }
\newcommand{\eg}{e.g., }

\journal{Artificial Intelligence}

\begin{document}

\begin{frontmatter}
%% TITLE MATTERS

\title{Beyond Incompatibility: Trade-offs between Mutually Exclusive Fairness Criteria in Machine Learning and Law}

\author[add1]{Meike~Zehlike\corref{cor1}}
\ead{zehlike@gmail.com}

\author[add1,add2]{Alex~Loosley}
\ead{aloosley@alumni.brown.edu}

\author[add3]{Håkan~Jonsson}
\ead{hakan.jonsson@zalando.de}

\author[add4]{Emil~Wiedemann}
\ead{emil.wiedemann@fau.de}

\author[add5]{Philipp~Hacker\corref{cor1}}
\ead{hacker@europa-uni.de}

\cortext[cor1]{Corresponding author}
\address[add1]{Zalando Research, Berlin, Germany}
\address[add2]{Echomotion GmbH, Munich, Germany}
\address[add3]{Zalando, Berlin, Germany}
\address[add4]{Department of Mathematics, Friedrich-Alexander-Universit\"at Erlangen-N\"urnberg, Erlangen, Germany}
\address[add5]{European New School of Digital Studies, European University Viadrina, Frankfurt (Oder), Germany}

\begin{abstract}
Fair and trustworthy AI is becoming ever more important in both machine learning and legal domains. One important consequence is that decision makers must seek to guarantee a `fair', i.e., non-discriminatory, algorithmic decision procedure. However, there are several competing notions of algorithmic fairness that have been shown to be mutually incompatible under realistic factual assumptions. This concerns, for example, the widely used fairness measures of ‘calibration within groups’ and ‘balance for the positive/negative class,’ \textcolor{maincolor}{which relate to accuracy, false negative and false positive rates, respectively}. %Indeed, the COMPAS algorithm, which predicts recidivism risk of criminal offenders, exhibits racial bias according to the balance metrics, but not regarding calibration.
In this paper, we present a novel algorithm (FAir Interpolation Method: FAIM) for continuously interpolating between these three fairness criteria. Thus, an initially unfair prediction can be remedied to meet, at least partially, a desired, weighted combination of the respective fairness conditions. %The algorithm relies on methods from the mathematical theory of optimal transport.
We demonstrate the effectiveness of our algorithm when applied to synthetic data, the COMPAS data set, and a new, real-world data set from the e-commerce sector. \textcolor{maincolor}{We provide guidance on using our algorithm in different high-stakes contexts, and} we discuss to what extent FAIM can be harnessed to comply with conflicting legal obligations. The analysis suggests that it may operationalize duties in traditional legal fields, such as credit scoring and criminal justice proceedings, but also for the latest AI regulations put forth in the EU, like the Digital Markets Act and the recently enacted AI Act. \end{abstract}
%\vspace{4pt}

	%\item Interpolation between mutually exclusive fairness criteria via optimal transport
	%\item Presentation of a novel method (FAIM) to reach any solution on the fairness simplex
	%\item Discussion on how FAIM can be harnessed to comply with conflicting legal obligations


\end{frontmatter}

%\newpage
\section{Introduction}

%test

Fairness and non-discrimination have increasingly moved to the center of AI research and policy in recent years \cite{mehrabi2021survey,ferrara2023fairness,zehlike2022fairness,pessach2023algorithmic,calo2017artificial,zuiderveen2018discrimination}. While studies on bias and discrimination perpetuated and exacerbated by computer systems have a long pedigree \cite{friedman1996bias,heinrichs2022discrimination}, recent advances in the AI space, including generative models, have brought an increased sense of urgency to this field \cite{bianchi2023easily,kotek2023gender,haim2024s,hacker2024generative}. As machine learning (ML) progressively permeates all sectors of society, responsible and fair AI development is more crucial than ever \cite{dignum2019responsible,grodzinsky2011developing,thiebes2021trustworthy}. This is particularly important in sensitive applications such as healthcare, hiring, credit, and law enforcement, where biased AI can have significant, real-world consequences.

Addressing fairness and non-discrimination has become a key concern for the AI community, not only but also for instrumental reasons: \textcolor{maincolor}{while furthering social justice, they also} enhance the robustness and generalizability of AI models. Biased models are likely to perform poorly when applied to certain subgroups or when faced with data that differs from their training set, limiting their applicability and effectiveness \cite{hendrycks2021many,alcover2023biased,grcic2024dense}. By prioritizing diversity and fairness, researchers can create AI systems that are more accurate and reliable across a variety of contexts and populations. This broad applicability seems essential for the long-term success and acceptance of AI technologies.

The devil, however, is in the details. For instance, in a much-noted incident, Google's Gemini model generated pictures of black and racially diverse US founding fathers, popes, and Nazi soldiers – in striking contrast to historical fact \cite{jacobi2024we}. This example highlights that it is often difficult, and sometimes even impossible, to enforce a unidimensional fairness metric for appropriate model behavior. Rather, in many settings, different varieties of fairness have to be traded off against one another, and against the desire for (historical) accuracy, performance, and other competing normative goals.

These inherent complexities and trade-offs will only become more important as specific legal provisions increasingly mandate fairness and non-discrimination in AI \cite{laux2022trustworthy,hacker2024generative,xenidis2019eu}. In this context, discrimination against protected groups in decisions facilitated by ML models remains a key challenge. Non-discrimination legislation applying to biased AI systems does exist in the US and the EU, but loopholes and enforcement problems remain \cite{barocas2016big,hacker2018teaching,zuiderveen2020strengthening,wachter2021fairness,wachter2022theory}. The new EU AI Act\footnote{\textcolor{maincolor}{Regulation (EU) 2024/1689 of the European Parliament and of the Council of 13 June 2024 laying down harmonised rules on artificial intelligence (Artificial Intelligence Act), OJ L, 2024/1689.}} compels AI developers to comprehensively assess and mitigate risks to fundamental rights, such as non-discrimination, stemming from high-risk AI systems and large general-purpose models (Art. 9, 55 AI Act) \cite{veale2021demystifying,ebers2021european,hacker2021legal}. This is backed up by a new framework for AI liability adopted by the EU legislator \cite{hacker2022european}.
Moreover, the EU Digital Markets Act (DMA\footnote{Regulation (EU) 2022/1925 of the European Parliament and of the Council of 14 September 2022 on contestable and fair markets in the digital sector, OJ 2022, L265/1.}) now compels large online platforms to engage in ‘fair and non-discriminatory’ ranking (Article 6(5) DMA), with a significant impact on the AI systems used for these tasks, which are key to the digital economy.

The laudable idea behind such approaches is to force developers to consider, and potentially eliminate, biases during the design of ML models.
Such non-discriminatory models are called ‘fair’ in the ML community \cite{mitchell2021,dwork2012awareness}.
However, foundational papers at the intersection of machine learning and the law have shown that, under realistic assumptions, several desirable fairness criteria are mutually incompatible \cite{kleinberg2016inherent}.
This poses a significant challenge as the law, in general, prohibits discrimination without explicitly favoring one metric over another.
Rather, a core tenet of legal research and jurisprudence is that conflicting rights or interests need to be balanced, i.e., be partially realized to satisfy each of the mutually incompatible demands to the extent that conditions warrant \cite{de2013balancing,morijn2006balancing}.
Lawmakers, courts and regulators therefore need fairness frameworks that move beyond the mere statement of incompatibility.
Simultaneously, ML research may benefit from the legal tradition of balancing by inducing decision makers and developers to refine algorithms implementing a balance between the normative trade-offs inherent in the mutually incompatible fairness criteria \cite{hertweck2022gradual}. Put succinctly: in many situations, implementing such trade-off algorithms will be a prerequisite for deploying AI in a legally compliant way at all. However, AI research may also benefit from integrating more complex boundary conditions in the long term, be it in generative AI or regression and classification tasks.
%Simultaneously, ML research may benefit from the balancing tradition of the law by developing and refining algorithms that enable decision makers to explicitly model, and resolve in situation-specific ways, the normative trade-offs inherent in mutually exclusive fairness criteria. \alex{This sentence confuses me.  It suggests that FAIM allows decision makers to model the normative trade-offs inherent in mutually exclusive fairness criteria. It also reads that decision makers can resolve normative trade-offs inherent in mutually exclusive fairness criteria.   Does that make sense?}

Our paper develops one such algorithm allowing decision makers to interpolate between three different incompatible fairness criteria singled out for their importance in previous research.
These three fairness criteria are described below.
There are various ways in which unfairness, i.e., discrimination against certain groups ~\cite{friedman1996bias}, may enter an ML model \cite{calders2013unbiased}.
For example, the model may perpetuate historical biases present in the training data; use non-representative training data; or optimize for a target variable which unwittingly encodes differential access of individuals or groups \cite{barocas2016big}.
The result of such biased modeling may be that false positive or false negative rates differ between groups of individuals.
Specific fairness metrics such as equalized odds, quantifying the difference of false positive and false negative rates between groups ~\cite{hardt2016equality}, are used to describe how much a classifier systematically rewards or punishes one group more than another with respect to ground truth.
Even if a classifier is fair in terms of equalized odds, its prediction scores can lead to unfair outcomes, however, if they are uncalibrated, as decision makers are not able to reliably interpret its output \cite{zadrozny2001obtaining, jiang2012calibrating}.\footnote{A classifier is calibrated when, given a score $s$, $s$-fraction of all individuals receiving score $s$ are ground truth positive \cite{dawid1982well}.}
What is worse, if a classifier emits scores that are differently calibrated between groups, identical scores have different meanings for different protected groups, again leading to unfair outcomes \cite{kleinberg2016inherent}.

Given that equality of false positive rates between groups, equality of false negative rates between groups, and calibration between groups are all important, socially and even legally desirable fairness criteria \cite{corbett2018measure}, it would be ideal if a classifier could satisfy all three.
Unfortunately, previous work shows that simultaneously fulfilling all three of these fairness criteria is
impossible, except under highly constrained circumstances such as when a classifier achieves 100\% accuracy with each instance
receiving a prediction score of exactly
0 or 1 \cite{chouldechova2017fair, kleinberg2016inherent, pleiss2017calibration}.
Merely focusing on one of these metrics may not do justice to the various interests and rights of the groups, and, as mentioned above, might even amount to unjustified discrimination if the law demands a balance rather than an absolute priority of one criterion.

Building on the assumption that decision makers should determine, within the constraints of the law, which combination of the three fairness criteria is most crucial towards a fair result in practice, our work introduces a new post-processing algorithm called FAir Interpolation Method (\methodname).
It allows decision makers to consciously and continuously interpolate between fairness criteria.
The algorithm is based on the notion that, given a classifier, a representative evaluation data set, and the
corresponding prediction scores for each group, any one of the three fairness criteria could be (partially)
satisfied by applying a certain fairness-criteria-specific score mapping function, which maps scores from the
distribution outputted by the model to a fair score distribution.
Given three such fairness-criteria-specific score distributions, \methodname uses optimal transport to
continuously interpolate between them. This leads to a mapping function that can be applied to new
classifier predictions, post hoc, to achieve a desired, weighted combination of the respective fairness criteria.
\methodname cannot be used to fully satisfy all three fairness criteria simultaneously,
but it does give the practitioner a powerful tool to make a continuous compromise along the fairness criteria simplex, and thus to account for different concerns, and legal obligations, each fairness criterion encompasses.

The remainder of the article is organized as follows.
First, related work is outlined in detail.
Next, the mathematical foundations regarding fairness criteria incompatibility and optimal transport are addressed.
Based on this, \methodname is introduced, and tested empirically in three experiments.
In the first experiment, \methodname is applied to a synthetic classifier that produces two distinct score distributions for
two respective groups.
Here the properties of \methodname are elucidated under controlled conditions.
In the second experiment, \methodname is tested on the COMPAS data set, on which the COMPAS algorithm famously scores individuals for recidivism risk.
In the last experiment, \methodname is applied to product rankings from the European e-commerce platform
Zalando.\footnote{\url{www.zalando.de}} Here, we evaluate our model in terms of typical fairness problems arising in e-commerce rankings.
The discussion takes up the results of the experiments and adds a specific legal perspective. We show how \methodname may be harnessed to trade off conflicting legal requirements across a wide variety of domains: credit scoring, criminal justice decisions, and fair rankings according to the DMA and the AI Act. \textcolor{maincolor}{The paper also provides guidance on how the different fairness criteria can be weighted and traded off against one another in a variety of high-stakes settings, ranging from recidivism prediction, credit scoring and e-commerce to hiring, college admissions, and healthcare.}

Overall, the paper brings together technical and legal perspectives on fairness criteria to take both discourses beyond incompatibility statements. In this way, decision makers may flexibly adapt ML models to different use cases in which varying interests of protected groups may necessitate different trade-offs between the involved fairness criteria. Simultaneously, gradually balancing the different fairness criteria, rather than prioritizing only one of them, will often be conducive to fulfilling legal requirements, both in established legal domains and in novel regulatory frameworks, such as the DMA and the AI Act.

\section{Related Work}

The incompatibility of various different fairness criteria in algorithmic decision making has been the subject of intense research in recent years. ~\citet{friedler2016impossibility} discuss the contrast between individual and group fairness. While these results hint at an incompatibility between individual and group fairness and relate these notions to two opposing worldviews, there is no rigorous proof of such incompatibility, and no discussion of possible intermediate worldviews. Some of the results in ~\citet{feldman2015certifying} and ~\citet{zehlike2020matching} can be viewed as proposals on how to continuously navigate between the two extremes.

The recidivism prediction problem was treated in ~\citet{kleinberg2016inherent, chouldechova2017fair}, with the result that a prediction algorithm must necessarily be unfair with regard to certain fairness criteria under realistic factual assumptions. In particular,~\citet{kleinberg2016inherent} show the mutual incompatibility between calibration within groups and balance for the positive/negative classes in the (typical) case of unequal base rates. These fairness criteria also underlie the present contribution. The framework of~\citet{kleinberg2016inherent} is thoroughly reviewed below; let us remark here only that the mentioned criteria of ~\citet{kleinberg2016inherent}~acknowledge and accept the possibility of empirically observed disparities between different groups, but require the equal treatment of individuals in different groups conditional on their similarity according to `ground truth'. The competing criterion of calibration within groups concerns the accuracy of the algorithm and is shown to conflict with the other two criteria.

\textcolor{maincolor}{While optimal transport provides a mathematically principled framework for interpolating between fairness criteria, it remains important to demonstrate why this approach was selected and how it compares to potential alternatives. In the current literature, there are few established methods for continuously navigating between different fairness notions. Most existing approaches treat fairness criteria as binary constraints that must be either fully satisfied or not, without allowing for meaningful intermediate solutions. The optimal transport framework, in contrast, provides a mathematically rigorous way to quantify and optimize the trade-offs between competing fairness criteria. By leveraging the theory of optimal transport, our approach guarantees that the interpolation between fairness criteria is optimal in terms of minimizing the cost of transforming one fairness regime into another, while preserving the underlying probabilistic structure of the problem\cite{zehlike2020matching}.}

\textcolor{maincolor}{Some previous work has looked into the possibility of simultaneously fulfilling several fairness constraints in modified ways. \citet{pleiss2017calibration} attempt to find a relaxed notion of error rate difference that can be	simultaneously achieved alongside group calibration.
However, they demonstrate that one may only simultaneously achieve group calibration with a relaxed version of one of either equalized false positive or false negative rates.
Furthermore, they show that classifiers simultaneously satisfying two of these three fairness criteria must engage in the equivalent of randomizing a certain percentage of predictions.
Thus, the authors suggest that practitioners should only attempt to satisfy one of the three fairness criteria based on importance, instead of trying to simultaneously fulfill a relaxed notion of two of them.
Often, however, all three fairness notions embody valid concerns of different groups \cite{corbett2018measure}. This is what FAIM acknowledges.}

\textcolor{maincolor}{One might consider alternative approaches such as quota-based systems, which have historically been used in domains like college admissions or hiring to balance competing fairness objectives. However, such methods suffer from several limitations. First, quota systems are typically applicable only to specific scenarios involving discrete allocation decisions, while our FAIM framework can be applied broadly across different types of algorithmic decision-making systems. Second, quota-based approaches often implement rather crude trade-offs, whereas our optimal transport method allows for more nuanced, continuous adjustments between different fairness criteria. Additionally, quota systems may face legal challenges in many jurisdictions, particularly under anti-discrimination laws in the US and the European Union, which generally prohibit rigid quota systems in favor of more flexible approaches to ensuring fairness \cite{hacker2018teaching,gee2024unprecedented}.}
\textcolor{maincolor}{A theoretical alternative might involve direct linear interpolation between different fairness metrics. However, such an approach would fail to account for the geometric structure of the probability spaces involved and could lead to implementations that violate basic probabilistic constraints. Our optimal transport-based framework, by contrast, explicitly preserves the probabilistic nature of the problem while providing a mathematically principled way to navigate the space of fairness criteria. This ensures that intermediate solutions remain well-defined probability distributions and maintain desirable statistical properties throughout the interpolation process.}

More generally, the use of optimal transport methods in algorithmic fairness has become increasingly popular ~\cite{dwork2012awareness, feldman2015certifying, gordaliza2019obtaining, zehlike2020matching, chiappa2020general, chiappa2021fairness}, applying optimal transport to trade-offs between individual vs group fairness, and accuracy vs fairness. Optimal transport is a mathematical tool that provides a very natural notion of distance of two probability distributions (the \emph{Wasserstein} or {earth mover distance}) and, moreover, gives an optimal (in a specific sense) way to translate between these distributions by means of an \emph{optimal transport map} or at least an \emph{optimal transport plan}, the latter requiring some form of randomization. In \cite{zehlike2020matching}, particularly, an important role is played by the \emph{Wasserstein-2 barycenter} of several probability distributions and the \emph{displacement interpolation} that allows to continuously interpolate between various distributions. We make use of these techniques in the present article.

The literature at the intersection of law and computer science has, in the past few years, increasingly discussed how legal non-discrimination requirements can be integrated into code, and vice versa ~\cite{zehlike2020matching,hacker2018teaching,wachter2021bias,wachter2021fairness,zuiderveen2020strengthening,gerards2022protected,wachter2020affinity,wachter2022theory,barocas2016big,raghavan2020mitigating,hellmann2020measuring,nachbar2020algorithmic}. This strand of research has focused particularly on remedying discrimination in criminal justice proceedings \cite{berk2021fairness,starr2014rationalization,mayson2019bias,selbst2017policing,katyal2019accountability,eaglin2017recidivism,huq2019equity,pruss2021distortion} and in other high-stakes AI decisions \cite{veale2018fairness}, such as employment contexts \cite{raghavan2020mitigating}, credit scoring \cite{bono2021algorithmic}, or university admissions \cite{zehlike2020matching}. More specifically, \citet{wachter2021bias} rightly point out that the reliance on fairness metrics based on ground truth may perpetuate biases if that ground truth itself is skewed against protected groups. This is a concern, for example, in criminal justice settings \cite{berk2021fairness,bao2021compaslicated}. We incorporate this constraint into the application perimeter of our algorithm (see Part~\ref{sec:discussion:compas and missing data}). Conversely, in the e-commerce sector, numerous publications deal with the emerging competition law framework for digital markets, for example the DMA \cite{eifert2021taming,hacker2022ki,brouwer2021towards,cabral2021eu,laux2021taming,podszun2021digital}. However, they do not, to our knowledge, specifically develop algorithms to solve legal questions arising in this field. The same holds true for the AI Act, where the literature, so far, focuses almost exclusively on the legal, economic and policy issues \cite{veale2021demystifying,laux2022trustworthy,hacker2023regulating,helberger2023chatgpt,novelli2024generative,schuett2023risk} or on explainability questions \cite{panigutti2023role,sovrano2022metrics}.

Existing publications often use US law as the relevant yardstick \cite{barocas2016big,raghavan2020mitigating,wachter2022theory,selbst2022unfair,hellmann2020measuring,gillis2019big,kim2017auditing,kim2022race}, but have lately increasingly incorporated explicit discussions of EU law as well~\cite{zehlike2020matching,hacker2018teaching,wachter2021bias,wachter2021fairness,zuiderveen2020strengthening,gerards2022protected,wachter2020affinity,wachter2022theory,veale2017fairer}. Our paper expands this strand of research by offering a novel method to trade off and implement competing technical and legal fairness constraints, and by discussing the EU Digital Markets Act (DMA) and AI Act in the context of technical algorithmic fairness. In three case studies, it demonstrates the potential of \methodname to integrate varying legal dimensions of fairness in ML settings, particularly in the criminal justice and credit scoring domain. The third case study tackles an, to our knowledge, entirely new field of law from an algorithmic fairness perspective: the DMA and the AI Act.

\section{Mathematical Theory and Algorithmic Implementation}
This section introduces \methodname, our interpolation framework that allows a continuous shift between three mutually exclusive fairness notions, as presented in~\citet{kleinberg2016inherent}.
We first provide a brief summary of the findings from~\citet{kleinberg2016inherent} to make the reader familiar with the three fairness notions and establish necessary terminology.
We then briefly present mathematical preliminaries from optimal transport theory.
Finally we describe the mathematical theory of our algorithm as well as its implementation in parallel, to help the reader translate between the formulas and the code.
We present the method in pseudocode in Algorithm~\ref{alg:fia}.

For the simplicity of presentation, we consider a population of individuals partitioned into two groups, indexed by $t=1,2$.
Note first that individuals need not be human, but could also be products, companies, etc., and second, that our method applies just as well to a setting with more than two groups.

Each individual, regardless of group membership, is either truly `positive' or `negative' with respect to some trait of interest; \ie in criminal justice, an individual would be `positive' if they were to commit a certain type of crime within a given period in the future, and negative if not.
The attributes `positive' and `negative' do not carry any normative meaning and are thus interchangeable, but we will refer to such a classification as `ground truth': It is assumed that the assignments of the positive/negative label reflect reality (we will shortly discuss the limitations of this assumption).

Furthermore, each individual is assigned a score by some algorithm (in the broadest sense).
The score, which can take as its value any real number in the interval $[0,1]$, is supposed to reflect the probability that the given individual is positive.
Therefore, if the ground truth were fully known to the algorithm, it would assign value 1 to positive individuals and 0 to negative ones.
This unlikely scenario is known as \emph{perfect prediction}.

A note on terminology: When we say that an individual is {\em truly negative/positive}, we mean that the individual is negative/positive according to ground truth, regardless of their predicted score. This should not be confused with the common terminology in which a `true negative' is an individual who is negative in ground truth {\em and} who, in addition, receives a negative prediction. In our setting, anyway, the predictions are not binary, but in terms of a continuous score between zero and one.

\subsection{Incompatibility of Fairness Criteria}
\label{sec:theory:kleinbergSummary}
The main result from \citet{kleinberg2016inherent} states that, in the setting described above, three natural fairness criteria are mutually incompatible except under trivial circumstances. We shortly summarize this incompatibility theorem here through a simple proof.
Eight quantities are of importance in this context:

\begin{itemize}
\item $N_t$ is the total number of individuals in group $t$;
\item $n_t$ is the number of (truly) {positive} individuals in group $t$;
\item $x_t$ is the average score of the individuals in group $t$ that are (truly) negative;
\item $y_t$ is the average score of the individuals in group $t$ that are (truly) positive.
\end{itemize}

For the purpose of deriving the incompatibility theorem, it need not be assumed that the ground-truth dependent quantities $n_t, x_t, y_t$ be known; rather, \emph{whatever their values might be}, the fairness criteria proposed in the next paragraph are never \textcolor{maincolor}{simultaneously} met except in case of perfect prediction or equal base rates.
Here, by \emph{base rate} for group $t$ we mean the ratio $n_t/N_t$, and \emph{perfect prediction} means that the score of each individual is 1 when she is positive and 0 when she is negative.

The fairness criteria are as follows:
\begin{itemize}
\item[A)] Calibration within groups: This measures the \emph{accuracy of prediction} of the score. The criterion requires that \emph{the average score in each group should equal the ratio of (truly) positive individuals in that group}, i.e.,
\begin{equation*}
\frac{x_t(N_t-n_t)+y_t n_t}{N_t}=\frac{n_t}{N_t},\quad t=1,2.
\end{equation*}
\item[B)] Balance for the negative class: \emph{The average score of (truly) negative individuals in group 1 should equal the average score of (truly) negative individuals in group 2}, i.e.,
\begin{equation*}
x_1=x_2=:x,
\end{equation*}
\textcolor{maincolor}{where the symbol `$=$:' denotes `equal by definition', i.e., $x$ is defined as the common value of $x_1$ and $x_2$ in the balanced case.}
\item[C)] Balance for the positive class: \emph{The average score of (truly) positive individuals in group 1 should equal the average score of (truly) positive individuals in group 2}, i.e.,
\begin{equation*}
y_1=y_2=:y.
\end{equation*}
\end{itemize}
Let us illustrate the three requirements for the case of recidivism prediction.
The members of each group (say, black and white offenders) are given a score that is intended to reflect the probability of the respective individual to recidivate within a given time period.
%
%A score close to 1 suggests a high probability of recidivism, while a score close to 0 suggests a low one.
%%
%A score of exactly one, for instance, suggests that the individual will commit an offense again with certainty.
Criterion~A) then requires the following: The \emph{average} score within group $t$ should equal the ratio of actual recidivists in that group; so if, say, half of the members of the white group would turn out recidivist, then the average score given to white individuals should be $0.5$.
%
%Note that this requirement is very weak, in that it says nothing about the validity of the score given to an \emph{individual}: In fact, in the example just given, even the algorithm that gets it completely wrong in assigning score 1 to the non-recidivists and zero to the recidivist would still produce the formally correct average of $0.5$, and thus pass requirement A). %
%Another quite useless algorithm would be the one that assigns score $0.5$ to each individual, thus containing no information on the individual level; yet criterion A) is still satisfied.
%%
%However, in the context of incompatibility, the weakness of A) does \emph{not} embody a weakness of the theory of~\cite{kleinberg2016inherent}; quite in contrast, an incompatibility result between certain requirements is \emph{stronger} the weaker the criteria are.
%
Concerning B), given a non-recidivist individual, it is desirable that their score be independent of their group membership. %Equivalently, the condition ensures that the rate of false positive predictions is the same for each group.
It would have to be considered highly unfair if a black person received a higher score, and thus faced a higher likelihood of detention, than a white person, given that both would actually not commit an offense if released.
This, in fact, is at the heart of the controversy around the COMPAS algorithm.
%\philipp{Please insert a sentence here saying the balancing the scores of true negative individuals between groups is equivalent to balancing the score of false positive individuals between groups.}
%
Note that B) only considers scores on an average level: The \emph{average} score among all truly non-recidivist individuals should be equal between groups.
Of course, this tells nothing about higher order statistical moments, such as the \textcolor{maincolor}{standard deviation} of the respective score distributions. %; for instance, it would allow for all white non-recidivists to receive a score of $0.2$, while many black non-recidivists obtain $0.1$ but a few get, say, $0.5$ -- as long as the average score in the black group is still $0.2$.
The discussion for criterion C) is analogous.

%\philipp{Please insert a sentence here saying the balancing the scores of true positive individuals between groups is equivalent to balancing the score of false negative individuals between groups.}

Suppose all the three requirements are satisfied, then this implies
\begin{equation}\label{fairness}
\begin{aligned}
x(N_1-n_1)+y n_1&=n_1,\\
x(N_2-n_2)+y n_2&=n_2.
\end{aligned}
\end{equation}
Regarding $N_t, n_t$ as given parameters, this is a linear system of two equations for the two unknowns $x,y$. It has a unique solution if and only if the determinant of the coefficient matrix is nonzero. This determinant is computed as
\begin{equation}\label{determinant}
(N_1-n_1)n_2-(N_2-n_2)n_1,
\end{equation}
which is zero if and only if either
\begin{inparaenum}[(1)]
	\item \textcolor{maincolor}{$n_1$ and $n_2$ are both not zero} and $\frac{N_1-n_1}{n_1}=\frac{N_2-n_2}{n_2}$, which is the case of equal base rates; or
	\item one of $n_1$ or $n_2$ is zero, in which case (assuming $N_1,N_2>0$) also the other one is zero, \ie there are no positive individuals at all.
	Here, the choice $x=0$ and $y$ arbitrary yields a fair assignment according to Eq.~\ref{fairness}. If there are no positive individuals, then everyone should get score zero, which is thus an instance of \emph{perfect prediction}.
\end{inparaenum}
For the first case, there are infinitely many solutions, one of which is $x=y=\frac{n_1}{N_1}=\frac{n_2}{N_2}$ (as mentioned in \citet{kleinberg2016inherent}).
This is realizable by assigning the same score to everyone. (A mathematically simpler solution would be $x=0, y=1$, which however requires the algorithm to know who is positive and who isn't and thus again requires perfect prediction.)

Finally, let's turn to the case when the determinant in Eq.~\ref{determinant} is nonzero, so there exists a unique solution to Eq.~\ref{fairness}.
It thus suffices to give one solution to rule out any other ones, and we can simply take $x=0, y=1$, \ie perfect prediction, which therefore is the only possibility.

In summary, we have thus verified Theorem 1.1, the main result of \citet{kleinberg2016inherent}, which we paraphrase as follows:
\begin{theorem}
For $t=1,2$, let $N_t, n_t>0$. If a score function satisfies A), B), and C), then
\begin{equation*}
\frac{n_1}{N_1}=\frac{n_2}{N_2}
\end{equation*}
(equal base rates), or the positive individuals receive score 1 and the negative individuals receive score 0 (perfect prediction).
\end{theorem}

\begin{remark}
The formulation given here of criterion A) is weaker than the one stated in \cite{kleinberg2016inherent} Page~4, where the criterion is required individually within each `bin'. As the computation here shows, the weaker requirement is already sufficient for incompatibility.
\end{remark}

\subsection{Mathematical Preliminaries on Optimal Transport}
Our goal is to design an algorithm that interpolates smoothly between the three fairness criteria discussed, as they are typically not attainable at the same time, \textcolor{maincolor}{as discussed in the preceding subsection}. The interpolation relies on the mathematical theory of optimal transport, which we now wish to recall.

A probability measure $\nu$ on $[0,1]$ is called {\emph{absolutely continuous}} if it is representable by an integrable probability density function $f:[0,1]\to \R$, which means that $\nu((a,b))=\int_a^b f(x)dx$ for any $a,b\in(0,1)$ with $a\leq b$. In words, the probability that a random variable distributed according to $\nu$ has its value in the interval $(a,b)$ is given by the integral of the density function $f$ over said interval. For this entire section, we will assume that all occurring probability measures are absolutely continuous. \textcolor{maincolor}{As our probability space $[0,1]$ is bounded, our measures will automatically have finite variance, meaning $\int_0^1 x^2 f(x)dx<\infty$}.
%\begin{assumption}\label{contassumption}
%The measures $\nu_1$ and $\nu_2$ are absolutely continuous and have finite variance, with integrable densities $f_1$ and $f_2$.
%\end{assumption}

Given two probability measures $\nu$ and $\mu$ on $[0,1]$, a \emph{transport map} $T:[0,1]\to[0,1]$ from $\nu$ to $\mu$ is a map such that, for every interval $(a,b)\subset[0,1]$,
\begin{equation*}
\mu((a,b))=\nu(T^{-1}(a,b)),
\end{equation*}
where $T^{-1}(a,b)$ denotes the preimage of $(a,b)$ under $T$, i.e., the set of all numbers that $T$ maps to the interval $(a,b)$. One often uses the notation $\mu=\nu\circ T^{-1}$ in this situation.
This means the following: If a random variable $X$ with values in $[0,1]$ is distributed according to $\nu$, then $T$ is a transport map from $\nu$ to $\mu$ if and only if the random variable $T(X)$ is distributed according to $\mu$.
As an example on $\R$, consider the two normal distributions $\mathcal N(0,1)$ and $\mathcal N(1,4)$, \textcolor{maincolor}{represented by Gaussian bell functions with expectation 0 and variance 1, or with expectation 1 and variance 4, respectively}. Then $T:\R\to\R$, $T(x)=2x+1$, would be a transport map from $\mathcal N(0,1)$ to $\mathcal N(1,4)$. However, there may be many transport maps: In this example, also $x\mapsto -2x+1$ would be a transport map.
Among the many transport maps, one (or potentially several) may be \emph{optimal} in the following sense:
\begin{definition}[Optimal transport map]
Let $\nu, \mu$ be two absolutely probability measures on $[0,1]$. An \emph{optimal transport map} is a transport map between $\nu$ and $\mu$ that minimises the cost functional
\begin{equation*}\label{cost}
C(\nu,\mu,T):=\int_0^1|x-T(x)|^2d\nu(x)
\end{equation*}
among all transport maps from $\nu$ to $\mu$.
\end{definition}
\noindent Under the stated assumptions on $\nu$ and $\mu$, ~\citet{brenier1987decomposition, brenier1991polar} ensures the existence and uniqueness\footnote{Uniqueness holds only $\nu$-almost everywhere; this means that the optimal transport map is not determined on any values outside the support of $\nu$.} of an optimal transport map. This allows to define the so-called quadratic \emph{Wasserstein distance} between $\nu$ and $\mu$, given by
\begin{equation*}\label{wasser}
W_2(\nu,\mu):= C(\nu,\mu,T)^{1/2}.
\end{equation*}
The Wasserstein distance forms a metric on the space of all absolutely continuous finite-variance probability measures on $[0,1]$.
The notion of optimal transport map allows to define a kind of continuous interpolation between the measures $\mu$ and $\nu$:
\begin{definition}[Displacement interpolation, cf.\ Remark 2.13 in~\cite{ambrosio2013user}]\label{displacement}
Let $\nu, \mu$ be two probability measures on $[0,1]$ with unique optimal transport map $T$. The \emph{displacement interpolation} between $\nu$ and $\mu$ with interpolation parameter $\theta\in[0,1]$ is defined by
\begin{equation*}
\nu^\theta((a,b)) = \nu((T^\theta)^{-1}(a,b)),
\end{equation*}
where the map $T^\theta:[0,1]\to[0,1]$ is given by $T^\theta(x)=(1-\theta) x+\theta T(x)$.
\end{definition}
\noindent Clearly, $\nu^0=\nu$ and $\nu^1=\mu$. The measure $\nu^\theta$ thus is a measure `in between' $\nu$ and $\mu$ that is closer to $\nu$ the smaller $\theta$ is chosen and closer to $\mu$ the larger it is. \textcolor{maincolor}{As an example, consider on the real numbers the normal distributions $\nu=\mathcal N(0,1)$ and $\mu=\mathcal N(1,1)$. The optimal transport maps from $\nu$ to $\mu$ in this case is $T(x)=x+1$, so that $T^\theta=x+\theta$, and correspondingly $\nu^\theta=\mathcal N(\theta,1)$. Note this is \emph{not} the same as the convex combination of the distributions, which would be $(1-\theta)\mathcal N(0,1)+\theta \mathcal N(1,1)$.} 
%For example, if (on $\R$ instead of $[0,1]$) $\nu=\mathcal{N}(0,1)$ and $\mu=\mathcal{N}(1,2)$, then
%\begin{equation*}
%\nu^\theta=\mathcal{N}(\theta,1+\theta).
%\end{equation*}

Finally, we mention the notion of \emph{barycenter} of a family of probability measures $\{\nu_k\}_{1,\ldots,N}$ with corresponding weights $\{w_k\}_{1,\ldots,N}$:

\begin{theorem}[Barycenter in Wasserstein space \cite{agueh2011bary}]
\label{bary}
Let\\ $\{\nu_k\}_{1,\ldots,N}$ be a family of absolutely continuous probability distributions, and let $\{w_k\}_{k=1,\ldots,N}$ be positive weights with $\sum_{k=1}^Nw_k=1$. Then there exists a unique probability measure $\nu$ on $[0,1]$ that minimizes the functional
\begin{equation*}
\nu\mapsto \sum_{k=1}^Nw_kW^2_2(\nu_k,\nu).
\end{equation*}
This measure is called the \emph{barycenter} of $\{\nu_k\}_{1,\ldots,N}$ with weights $\{w_k\}_{1,\ldots,N}$.
\end{theorem}
\noindent \textcolor{maincolor}{Thus, the barycenter gives the least square minimization on the level of probability measures, so to speak.} With two groups and respective score distributions $\nu_1, \nu_2$ and weights $w_1, w_2$, the barycenter is given by the displacement interpolation $\nu = \nu_1\circ (T^{w_2})^{-1}$, where $T$ is the optimal transport map between $\nu_1$ and $\nu_2$ and $T^{w_2}$ is given in Def.~\ref{displacement}. \textcolor{maincolor}{Specifically, in our example with $\nu=\mathcal N(0,1)$ and $\mu= \mathcal N(1,1)$ and weights, say, $w_1=0.2$ and $w_2=0.8$, the barycenter would be $\mathcal N(0.8,1)$.}

\subsection{The FAir Interpolation Method (\methodname)}
\label{sec:theory:algorithm}
The goal of this paper is to transform a given model output into one which `balances' between the three fairness criteria to an extent that the user is free to choose according to her specific situation.
For instance, a model (like COMPAS) that has been used for a while and has been evaluated to give quite accurate predictions, but discriminate unfairly between groups, can be transformed into an improved algorithm that respects group fairness to a higher degree.

Before we delve into the mathematics, let us describe more clearly the situation.
Again we have two groups of individuals represented as (disjoint) sets $X_1, X_2$, and for each group a given scoring algorithm, which for group $t$ ($t=1,2$) is simply a map $S_t: X_t\to[0,1]$.
We assume this algorithm has been used for a sufficient amount of time such that reasonably reliable data is available on its performance; more precisely, we assume for each group a map $[0,1]\to[0,1]$, $s\mapsto\lambda^+_t(s)$ to be given, where $\lambda^+_t(s)$ reflects the proportion of truly positive\footnote{Recall that `truly positive' means positive according to ground truth, regardless of the score received.} individuals from group $t$ that were assigned score $s$.
Alternatively, one may interpret $\lambda^+_t(s)$ as the probability that a randomly chosen individual from group $t$ whose score equals $s$ will be truly positive.
We also define $\lambda^-_t(s):=1-\lambda^+_t(s)$ for all $s\in[0,1]$ as the proportion of truly negative instances among individuals of group $t$ and score $s$.
From historical data, we know the score distributions in each group assigned by $S_t$, that is, we have two probability measures $\nu_1,\nu_2$ such that for any (measurable) set $Q\subset[0,1]$, $\nu_t(Q)$ is the probability that an individual $x$ randomly chosen from group $t$ has a score in $Q$.

\begin{remark}
In practice, the question arises how to collect the data encoded in $\lambda_t^+$ and $\nu_t$.
For the latter, it suffices to collect the previous outcomes of the scoring algorithm $S_t$, since $\nu_t(s)$ is the proportion of individuals from group $t$ that were assigned score $s$ in the past.
The integral of $\nu_t(s)$ over all possible scores will then equal one, because the probability that a given individual was assigned \emph{some} score is one.
For $\lambda_t^+$, we need information on the \emph{ground truth}, which is always a difficult and controversial matter.
In addition to the data on past scores, we need data on the actual positivity or negativity of individuals.
%
%For instance, one would need to know if a specific white person rated with a certain score by the COMPAS algorithm five years ago did become recidivist since then.
%
Though such data may be difficult to collect reliably, we wish to remark that such information is not just required for our method, but for \emph{any} reasonable quality control of the algorithm in question.
Without any (at least approximate) determination of ground truth, there is generally no way to evaluate the performance of a given algorithm.

%In any case, once such data from previously scored individuals is available, it is possible to divide the number of truly positive individuals in group $t$ that received score $s$ by the total number of individuals in group $t$ that received score $s$, and this gives the value of $\lambda_t^+(s)$.
\end{remark}

In the first steps of Algorithm~\ref{alg:fia}, we give for each of the criteria A), B), C) a procedure to transform a given algorithm that does not satisfy the respective criterion into one that does.

\subsubsection{Criterion A)}
The original algorithm might not yet be correctly calibrated, \ie the scores produced by $S_t$ do not yet give an accurate representation of the actual probability of being positive.
To fix it, we compute a new map $S_t^A$  defined by
\begin{equation}\label{eq:theory:criterionA:SA}
S_t^A(x):=\lambda^+_t(S_t(x)),
\end{equation}
which means that an individual from group $t$ who initially received score $s$ will now receive score $\lambda^+_t(s)$, which is precisely its probability of being positive.
In Algorithm~\ref{alg:fia}, Lines~\ref{alg:fia:sAStart}--\ref{alg:fia:sAEnd} compute map $S_t^A$, and Lines~\ref{alg:fia:muAStart}--\ref{alg:fia:muAEnd} yield the probability distributions $\mu^A_t$.
This probability, as discussed, is computed from historical data, which approximately reflects ground truth.
Thus, although the new score will not in general be able to give a deterministic value (zero or one) of the particular individual's true positivity, the new algorithm does satisfy requirement A) (even bin-wise, \ie for each $s$).

\subsubsection{Criterion B)} To express the balance criterion, we need to determine the expected score of an individual $x$ from group $t$ on the condition that $x$ is negative.
These conditional expectations for both groups should then be equal.
Indeed, we can find a new score map such that the entire probability distributions of the score, conditional on negativity, coincide, thus giving statistical parity in score between the negative instances in each group. In particular, the risk of a false positive outcome will be equal for both groups.

\begin{center}
\begin{algorithm}[H]
	\caption{Algorithm \methodname. First, criteria $A, B$ and $C$ are individually fully satisfied, yielding distributions $\mu^A_t, \mu^B_t$, and $\mu^C_t$. Then, the barycenter $\bar{\mu}_t$ of $\mu^A_t, \mu^B_t$, and $\mu^C_t$ is computed, which yields optimal transport map $T_t$. $T_t$ is finally used to map from the original score distribution $\nu_t$ to the fair distribution~$\bar{\mu}_t$. ``emd'' stands for earth mover distance.}
	\label{alg:fia}
	\AlgInput{
		$\texttt{rawScores[]}:$ an array with a score for each individual;\\
		$\texttt{stepsize}:$ a float that specifies the bin width of the truncated scores;\\
		$\texttt{groundTruthLabels[]}:$ an array with a binary ground truth label for each individual;\\
		$\texttt{groups[]}:$ an array with a group membership label for each individual; \\
		$\texttt{thetas[[]]}:$ a matrix with rows $\left[\theta^A, \theta^B, \theta^C\right]$ for each group $t$.}
	\AlgOutput{
		$\texttt{fairScores[]}:$ a new score array with a fair score for each individual such that the fair score distributions map the fairness criteria given by $\texttt{thetas}$.}

	\tcp{Normalize scores to $[0, 1]$ and discretize}
	$\texttt{rawScores[]} = \texttt{truncateScores(0, 1, stepsize)}$\\
%	\tcp{Initialize score maps}
%	\texttt{$S^A_t$ = [], $S^B$ = [], $S^C$ = []}\\
%	\tcp{Initialize arrays to contain score distributions, which meet the respective criterion}
%	\texttt{$\mu^A_t$ = [], $\mu^B$ = [], $\mu^C$ = []}\\
	\tcp{Step 1: Compute score map $S^A_t$, yielding $\mu^A_t$ (Eq.\ref{eq:theory:criterionA:SA})}
	\texttt{criterionAScores[] = zeros(rawScores[].length)}\\
	\For{\texttt{$t$ in groups[].uniqueValues()}}{
		\For{\texttt{$s$ in rawScores[].uniqueValues()}}{\label{alg:fia:sAStart}
			\texttt{groupMask = groups[] == $t$; groundTruthPositivesMask = groundTruthLabels[] == 1; scoreMask = rawScores[] == $s$;}\\
			\texttt{$s^A_t$ = $\frac{\texttt{rawScores[groupMask \&\& scoreMask \&\&  groundTruthPositivesMask].length}}{\texttt{rawScores[groupMask \&\& scoreMask].length}}$}\\
			\tcp{add key value pair to score map for criterion $A$.}
			$S^A_t$\texttt{.add(($s, s^A_t$))}
		}\label{alg:fia:sAEnd}
		\tcp{translate raw scores into fair scores w.r.t. criterion $A$: everybody with raw score $s$ gets fair score $s^A_t$ assigned.}
		\texttt{criterionAScores[groupMask] = rawScores[groupMask].translate($S^A_t$)}\label{alg:fia:muAStart} \\
		$\mu^A_t$ \texttt{ = criterionAScores[groupMask].histogram()} \label{alg:fia:muAEnd}\\

	}
	\tcp{Step 2: compute $\mu^B_t$ }
	\texttt{groupSigmas = [[]]}; \label{alg:fia:sigmaMinusStart} \texttt{groupSizesInPercent = []};\\
	\texttt{groundTruthNegativesMask = groundTruthLabels[] == 0}\\
	\For{\texttt{$t$ in groups[].uniqueValues()}}{
		\texttt{groupMask = groups[] == $t$}\\
		\texttt{groupSizesInPercent.add(groupMask.length() / groups[].length())}\\
		\tcp{compute Eq.~\ref{eq:theory:criterionB:sigmaPerGroup}}
		\texttt{$\sigma^-_t$ = zeros(rawScores.histogram().length())}\\
		\For{\texttt{$s$ in rawScores[].uniqueValues()}}{
			\texttt{scoreMask = rawScores[] == $s$}\\
			\texttt{percentGroundTruthNegative = $\frac{\texttt{rawScores[groupMask \&\& groundTruthNegativesMask \&\& scoreMask].length}}{\texttt{rawScores[groupMask \&\& groundTruthNegativesMask].length}}$}\\
			\texttt{$\sigma^-_t$.add(percentGroundTruthNegative)}
		}
		\texttt{groupSigmas.add($\sigma^-_t$)}\label{alg:fia:sigmaMinusEnd}
	}
	\tcp{compute barycenter $\bar{\sigma}^-$ between \texttt{groupSigmas}}
	$\bar{\sigma}^-$ \texttt{ = wasserstein2barycenter(groupSigmas, groupSizesInPercent)} \label{alg:fia:sigmaBar}\\

\end{algorithm}
\end{center}

\begin{algorithm}[t]
	\setcounter{AlgoLine}{25}
		\For{\texttt{$t$ in groups[].uniqueValues()}}{
		\texttt{groupMask = groups[] == $t$}\\
		\tcp{use $\bar{\sigma}^-$ to get optimal transport maps for ground truth negative individuals (Eq.~\ref{eq:theory:criterionB:transportMapsB})}
		$T^-_t$ \texttt{= computeOptimalTransportMap(rawScores[groupMask \&\& groundTruthNegativesMask].histogram(),  $\bar{\sigma}^-$, kind=emd)} \label{alg:fia:fairTransportMapCondB}\\
		\tcp{For ground truth negatives, translate \texttt{rawScores} into $S^B_t$. The histogram of $S^B_t$ forms $\mu_t^B$. Note, that $\mu^B_t$ is the score distribution for group~$t$ that contains \textbf{all} individuals (negatives and positive), even though the translation is done for true negatives only.}
		\texttt{$\mu_t^B$ = rawScores[groupMask \&\& groundTruthNegativesMask].translate($T^-_t$).histogram()} \label{alg:fia:muB}
	}
	\tcp{Step 3: compute $\mu^C_t$. We omit this part of the algorithm for brevity, since this is done the same way as $\mu^B_t$ (Lines~\ref{alg:fia:sigmaMinusStart}--\ref{alg:fia:muB}). The only difference is that we would use a groundTruthPositivesMask = groundTruthLabels[] == 1, instead of a groundTruthNegativesMask.}
	\tcp{Step 4: compute final barycenter $\bar{\mu}_t$ which incorporates thetas[[]] and optimal transport maps $T_t$ for each group $t$. Then translate rawScores of group into fair scores.}
	\For{\texttt{$t$ in groups[].uniqueValues()}}{
		\tcp{compute $\bar{\mu}_t$ (Eq.~\ref{eq:theory:finalBary})}
		\texttt{$\bar{\mu}_t$ = wasserstein2barycenter($\mu^A_t, \mu^B_t, \mu^C_t$, thetas[$t$])} \label{alg:fia:finalBary}\\
		\texttt{groupMask = groups[] == $t$}\\
		$T_t$ \texttt{= computeOptimalTransportMap(rawScores[groupMask].histogram(),  $\bar{\mu}_t$, kind=emd)} \label{alg:fia:fairTransportMap}\\
		\texttt{fairScores[groupMask] = rawScores[groupMask].translate($T_t$)} \label{alg:fia:fairScoreTranslation}
	}
	\Return{\texttt{fairScores[]} }
\end{algorithm}

We do this as follows (see again Algorithm~\ref{alg:fia}):
First (Lines~\ref{alg:fia:sigmaMinusStart}--\ref{alg:fia:sigmaMinusEnd}), we compute said conditional probabilities $\lambda_t^-$ and $\nu_t$ from the given data.
By Bayes' Theorem \textcolor{maincolor}{for continuous random variables}, the probability distribution for the score of an individual $x$ from group $t$ conditional on $x$ being negative is given as
\begin{equation}\label{eq:theory:criterionB:sigmaPerGroup}
\sigma^-_t:=\frac{\lambda_t^-(s)\nu_t(ds)}{\int_0^1 \lambda_t^-(s)d\nu_t(s)}.
\end{equation}
%In more concrete terms, if $\nu_t$ is given by the probability density function $f_t$, then the probability density function $g^-_t$ of $\sigma_t^-$ is given as
%\begin{equation*}
%g_t^-(s)=\frac{\lambda_t^-(s)f_t(s)}{1-\frac{n_t}{N_t}}.
%\end{equation*}
%(Note that the denominator is nothing but the base rate of negative instances in the group $t$.)
%
Secondly, we consider the Wasserstein-2 barycenter of $\sigma^-_1, \sigma^-_2$ on $[0,1]$ with weights $\frac{N_1}{N_1+N_2}$ and $\frac{N_2}{N_1+N_2}$, respectively, which we denote by $\bar\sigma^-$ (Line~\ref{alg:fia:sigmaBar}).
Then there are two (optimal) transport maps $T^-_t:[0,1]\to[0,1]$ \textcolor{maincolor}{between $\sigma^-_1$ and $\bar\sigma^-$, and between $\sigma^-_2$ and $\bar\sigma^-$,} such that
\begin{equation}
	\label{eq:theory:criterionB:transportMapsB}
\bar\sigma^-=\sigma^-_1\circ(T^-_1)^{-1}=\sigma^-_2\circ(T^-_2)^{-1}.
\end{equation}
Finally, set $S_t^B(x):=T^-_t(S_t(x))$ for $t=1,2$. Replacing $\nu_t$ with $\nu_t^B:=\nu_t\circ (T^-_t)^{-1}$, which is the score distribution under the modified algorithm $S_t^B$, and replacing also $\lambda^-_t$ with $(\lambda^-_t)^B:=\lambda_t^-\circ (T^-_t)^{-1}$, which is the new probability that an individual from group $t$ with modified score is negative, we obtain via Eq.~\ref{eq:theory:criterionB:sigmaPerGroup} (Line~\ref{alg:fia:muB})
\begin{equation}\label{eq:theory:conditionB:muB}
\mu_t^B:=\sigma^-_t\circ (T_t^-)^{-1}=\bar\sigma^-
\end{equation}
as the probability distribution of the modified score conditional on being negative.
Since they agree for both groups, the new $S_t^B$ satisfies (a much stronger version of) B).

Let us, however, not ignore a small issue here: The map $T^-_t$ is, as mentioned above, only well-defined on the support of $\sigma_t^-$. For our intended application, there might well be scores that have never been given to a truly negative (or positive) individual: For example, if the original algorithm was not completely misguided, then it would most probably never have given score 1 to a truly negative individual, so that $\sigma_t^-$ would vanish near 1. For such scores we are essentially free to define the map $T$. The most natural and simple choice is to leave these scores unchanged, \ie to set $T_t^-(x)=x$ for any $x$ outside the support of $\sigma_t^-$.

%A pair of algorithms $S_1^B, S_2^B$ therefore satisfies B) if and only if
%\begin{equation}
%\frac{\int_0^1 s\lambda_1^-(s)d\nu_1(s)}{\int_0^1 \lambda_1^-(s)d\nu_1(s)}=\frac{\int_0^1 s\lambda_2^-(s)d\nu_2(s)}{\int_0^1 \lambda_2^-(s)d\nu_2(s)}.
%\end{equation}

\subsubsection{Criterion C)} Replacing $-$ by $+$ everywhere in B), we obtain a new algorithm $S_t^C$ that satisfies C) and in particular assimilates the probabilities of a false negative outcome of the evaluation for both groups.

\subsubsection{Combining the three procedures}
Each of the modified score maps $S_t^A, S_t^B, S_t^C$ gives rise to a corresponding score distribution $\mu_t^A, \mu_t^B, \mu_t^C$.
For each group $t=1,2$, we therefore obtain a triangle in Wasserstein space, \textcolor{maincolor}{i.e., in the space of probability measures}.
Let now $\theta_t^A, \theta_t^B, \theta_t^C\in[0,1]$ be three real parameters that add up to $1$, which are given as input to the algorithm.
We then consider the weighted barycenter $\bar\mu_t$ in Wasserstein-2 space, \ie the unique probability measure $\bar\mu$ that minimizes the functional
\begin{equation}\label{eq:theory:finalBary}
\mu\mapsto\theta_t^A W_2^2(\mu,\mu_t^A)+\theta_t^B W_2^2(\mu,\mu_t^B)+\theta_t^C W_2^2(\mu,\mu_t^C).
\end{equation}
This barycenter embodies a `compromise' between our three incompatible fairness criteria, where the parameters $\theta = \theta_t^A, \theta_t^B, \theta_t^C$ allow to continuously adjust the weights that the decision-maker wishes to put on the respective criterion (Line~\ref{alg:fia:finalBary}). For this reason, $\theta$ are referred to below as fairness objective parameters.

The fair score of an individual in group $t$, based on the original (`unfair') score function $S_t$, is then determined as $T(S(x))$, where $T:[0,1]\to[0,1]$ is the optimal transport map from the original score distribution $\nu_t$ to the fair distribution $\bar\mu_t$ (Lines~\ref{alg:fia:fairTransportMap}--\ref{alg:fia:fairScoreTranslation}).

A final remark: By construction, whenever we optimize for only one of the fairness criteria (that is, one of the $\theta$'s is set to one), this respective criterion will be exactly satisfied, whereas the other two might actually get worse. An interesting mathematical question is whether, when the three criteria are `mixed' (i.e., none of the $\theta$'s is zero), it is possible that \emph{all three} fairness goals simultaneously deteriorate -- a sort of lose-lose situation. While we cannot exclude that this happens in pathological cases, it is certainly not an expected effect in realistic circumstances, and we have not observed such a lose-lose scenario in our experiments to be discussed in the next section.

\FloatBarrier

\section{Experiments}
We conduct extensive experiments on three use cases (one with synthetic data and two with real-world data) to show the effectiveness of \methodname to interpolate between the three different fairness criteria described in Section~\ref{sec:theory:kleinbergSummary}.\footnote{Data and code for all experiments are available under \url{https://github.com/MilkaLichtblau/faim}. The repository also contains clear documentation for using \methodname on one's own data.}
First, we apply \methodname to synthetic data with two demographic groups, one advantaged and one disadvantaged.
Second, we apply the same analysis to the \compas data set, which contains protected group variables for gender, race, and age.
We are aware of critical voices such as~\cite{bao2021compaslicated} on using \compas to benchmark new fairness methods, and discuss whether recidivism risk is an appropriate use case for \methodname in Section~\ref{sec:discussion:compas and missing data}.
However, since ~\citet{kleinberg2016inherent}, whose terminology we adopt, build their argumentation on the \compas case, we want to show how our method and findings compare to those of ~\citet{kleinberg2016inherent}.
Last, we study how the application of \methodname affects rankings on the European e-commerce platform Zalando, and whether it can be used to overcome the problem of popularity bias \cite{bellogin2017statistical}.
For each use case, \methodname is tested with four fairness objectives by setting the following combinations of the fairness objective parameters $\theta_t^A$, $\theta_t^B$ and $\theta_t^C$ (see Eq.~\ref{eq:theory:finalBary}):
\begin{enumerate}
\item Fairness criterion A: Calibration within groups ($\theta_t^A = 1, \:\forall t$)
\item Fairness criterion B: Balance for the negative class ($\theta_t^B = 1, \:\forall t$)
\item Fairness criterion C: Balance for the positive class ($\theta_t^C = 1, \:\forall t$)
\item An equally weighted combination of fairness criteria A), B), and C) ($\theta_t^A = \theta_t^B = \theta_t^C = \frac{1}{3}, \:\forall t$)
\end{enumerate}

Note that the $\theta$-array can be configured differently for each group $t$. However, for simplicity of presentation, we set the same $\theta$s for each group and drop subscript $t$.
For each fairness objective, we calculate performance metrics (accuracy, weighted average precision, and weighted average recall) and error rates including false negative rate (FNR) and false positive rate (FPR). Results from using \methodname are compared to those produced by the base model.

\subsection{Experiments on Synthetic Data}
As a motivating example for the synthetic data experiment, let us consider credit scoring. Increasingly, scoring agencies are turning to ML to predict creditworthiness \cite{langenbucher2020responsible,langenbucher2021responsible}. Now imagine that a company predicts ML-based credit scores for candidates stemming from two protected groups only, for example for men and women (no non-binary persons applied). Candidates receiving a score equal or above 0 are labeled offered a credit contract, while those below are rejected.

\subsubsection{Data set.}

To model such a credit scoring situation, we generate a large synthetic data sets with 100,000 individuals and two demographic groups (one advantaged, e.g., male; and one disadvantaged, e.g., female) of approximately the same size.
Each individual is assigned two scores, a \emph{true score}, and a \emph{predicted score}.
The true score is to be understood as an individual's true probability of being part of a desirable positive class (e.g., creditworthy person), while the predicted score is meant to represent a score that has been assigned to an individual by a (synthetic) scoring algorithm.
They are sampled from a multivariate normal distribution with covariance matrix $[[1, 0.8], [0.8, 1]]$ and true score means $(1, -1)$ for the advantaged and disadvantaged groups, respectively.
The corresponding predicted score means are $(2, -3)$.

\begin{wrapfigure}{r}{.5\textwidth}
	\vspace{0mm}
	\centering
	\includegraphics[width=.5\textwidth]{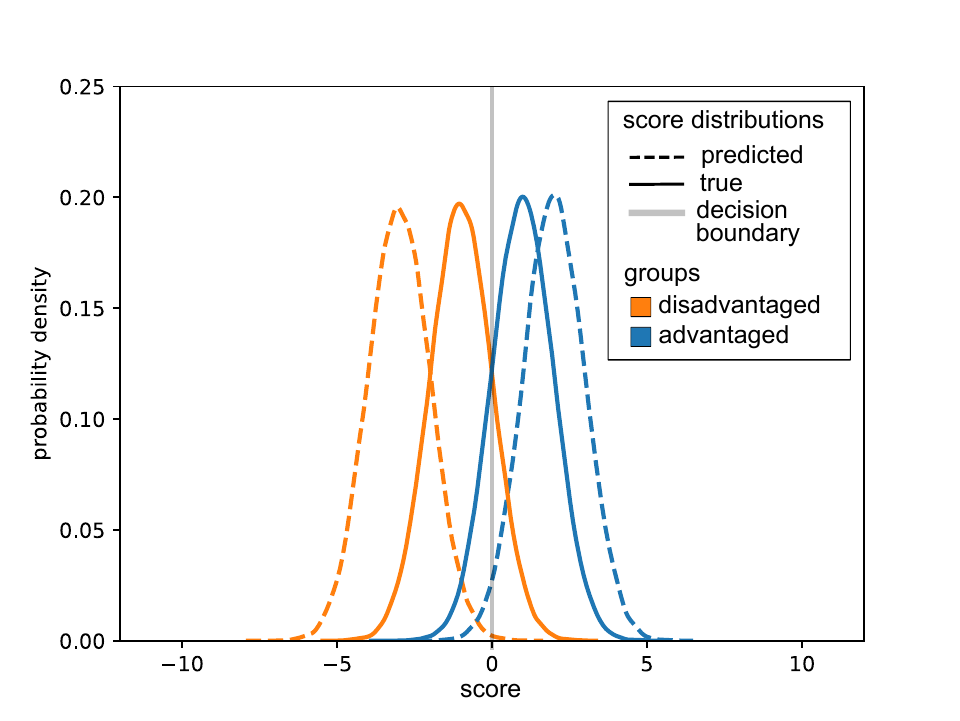}
	\caption{(Best seen in color) Synthetic true (solid lines) and predicted (dashed lines) score distributions for demographic two groups (orange and blue).  The decision boundary at $\text{score}=0$ determines ground truth positive and negative labels based on the true scores, and predicted positive and negative labels  based on the predicted scores. This simulates a situation in which the scoring algorithm overestimates the qualification of the advantaged group, and further disadvantages the disadvantaged group.}
	\vspace{-3mm}
	\label{fig:experiments:data sets:synthetic}
\end{wrapfigure}

Additionally, we define a synthetic decision threshold to yield ground truth and predicted labels: those with a true (resp. predicted) score above zero are considered ground truth (resp. predicted) positive (i.e., creditworthy, and receive a credit offer).
Conversely, those with a true (resp. predicted) score below zero are considered ground truth (predicted) negative (i.e., not creditworthy, and are rejected).
Fig.~\ref{fig:experiments:data sets:synthetic} depicts this data set.
Blue lines correspond to the advantaged group, with both predicted scores (dashed blue line) and true scores (solid blue line) normally distributed with peaks greater than zero.
Orange lines correspond to the disadvantaged group, with both predicted scores (dashed orange line) and true scores (solid orange line) normally distributed with peaks less than zero.
This data set reflects a scenario of different base rates between the groups, as the advantaged group has a higher true probability of being positive. In our example, this is the male group.
However, the scoring algorithm has overestimated the capabilities of the blue, advantaged group (the predicted score distribution is centered to the right of the true score distribution), and underestimated the capabilities of the orange, disadvantaged group (the predicted score distribution is centered to the left of the true score distribution).
Additionally, the algorithm does a better job in predicting the true score distribution for the advantaged group, than it does for the disadvantaged group (the blue distributions overlap more).
To emphasize the disparity between the two demographic groups, the advantaged group is 5.3 times more likely to be labeled positive based on the true scores, compared to 6250 times more likely to be labeled positive based on the predicted scores from the (unfair) scoring algorithm.

These two particular properties of disparities in model quality across groups are well studied in the algorithmic fairness literature~\cite{angwin2016machine, zafar2017fairness, hardt2016equality}, which is why we choose this synthetic setting to present the functioning of \methodname.
Note that we have plotted the true score distributions only for demonstration purposes to give a clear picture of the discriminatory scenario.
They are usually not available, which is why \methodname relies on \emph{observable} ground truth labels, as explained in Section~\ref{sec:theory:algorithm}.

\subsubsection{Results.}
\label{sec:experiments:synthetic:results}
\begin{figure}[b!]
	\vspace{0mm}
	\centering
	\includegraphics[width=\textwidth]{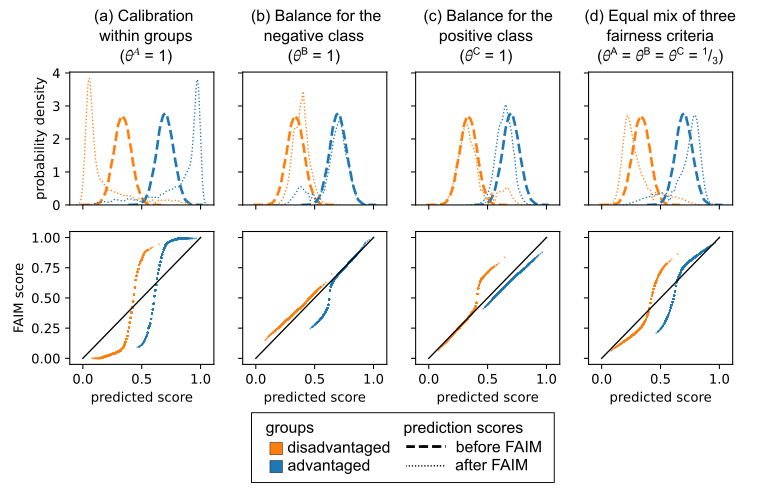}
	\caption{(Best seen in color). Prediction scores before and after applying \methodname to shift scores to achieve (a-c) each of the three fairness criteria, respectively, and (d) an equal mix of all three fairness criteria. The prediction scores before \methodname are the same as in Fig.~\ref{fig:experiments:data sets:synthetic} but scaled to values between 0 and 1. The score distributions (top row) have been smoothed using kernel density estimation to make comparison between distributions easier to interpret. The discrete score transport plots (bottom row) shows how mixing of fairness criteria (d) yields a score transport somewhere in between that corresponding to each respective fairness critera.}
	\vspace{-3mm}
	\label{fig:experiments:results:synthetic}
\end{figure}

Fig.~\ref{fig:experiments:results:synthetic} shows score distributions (top row) and corresponding discrete transport maps (bottom row) showing the changes in scores before and after applying \methodname with the four fairness objectives described above.
The transport maps are to be read as follows: each raw score on the x-axis gets replaced by a new score on the y-axis.
Note that the transport maps necessarily differ by group, since the groups experience disparate treatment by our synthetic model which underestimated the capabilities of the disadvantaged group, and overestimated those of the advantaged.
When the fairness objective was calibration within groups (Fig.~\ref{fig:experiments:results:synthetic}a), scores became polarized because the fraction of ground truth negative and positive for low and high scores was even lower and higher, respectively.
For the fairness objectives balance of the negative and positive classes (Fig.~\ref{fig:experiments:results:synthetic}b and \ref{fig:experiments:results:synthetic}c, respectively), score transport only applies to the subset of individuals that are ground truth negative and positive, respectively. Furthermore, the barycenter towards which scores are transported is weighted by the fraction of ground truth negative and positive in each group, respectively (see Eq.~\ref{eq:theory:criterionB:transportMapsB}).
Thus, for the balance of the negative class fairness objective, because the disadvantaged ground had a larger fraction of ground truth negative individuals, \methodname decreased the scores of advantaged ground truth negative individuals more than it increased the scores of disadvantaged ground truth negative individuals, thus giving the result in Fig.~\ref{fig:experiments:results:synthetic}b.
Similar logic can be applied to explain why high scores from the disadvantaged ground were mapped favorably when applying \methodname for balance of the positive class (Fig.~\ref{fig:experiments:results:synthetic}c).
Finally, when applying \methodname with an objective of an equal mix of the three fairness criteria, the scores and transport map in Fig.~\ref{fig:experiments:results:synthetic}d indeed show a interpolated compromise between the scores and transport maps in Fig.~\ref{fig:experiments:results:synthetic}a-c.

\setlength{\tabcolsep}{1pt}
\begingroup
\begin{table}[b!]
\footnotesize
\vspace{-5mm}
\begin{tabular*}{\textwidth}{@{\extracolsep{\fill}}*{6}{llllll}}
\toprule
\begin{tabular}{@{}c@{}}Synthetic \\ Data Set\end{tabular} & \multicolumn{3}{c}{Performance} & \multicolumn{2}{c}{Error Rates}   \\
\cline{2-4} \cline{5-6}
      & Accur. ($\Delta$)  & \begin{tabular}{@{}c@{}}Weighted Avg. \\ Precision ($\Delta$)\end{tabular} & \begin{tabular}{@{}c@{}}Weighted Avg. \\ Recall ($\Delta$)\end{tabular}	& FPR ($\Delta$) & FNR ($\Delta$) \\    \midrule
      before \methodname & 0.852 & 0.853 & 0.852 & 0.138 & 0.157 \\
      $\;\;\;$ blue & 0.860 & 0.868 & 0.860 & 0.869 & 0.002 \\
      $\;\;\;$ orange & 0.844 & 0.869 & 0.844 & 0.000 & 0.990 \\
      \midrule
      $\theta^A = 1$ & 0.885 (\textcolor{forestgreen}{0.033}) & 0.885 (\textcolor{forestgreen}{0.033}) & 0.884 (\textcolor{forestgreen}{0.032}) & 0.116 (\textcolor{forestgreen}{-0.022}) & 0.114 (\textcolor{forestgreen}{-0.043})  \\
      $\;\;\;$ blue & 0.884 (\textcolor{forestgreen}{0.024}) & 0.873 (\textcolor{forestgreen}{0.005}) & 0.884 (\textcolor{forestgreen}{0.024}) & 0.560 (\textcolor{forestgreen}{-0.309}) & 0.032 (\textcolor{red}{0.032})  \\
      $\;\;\;$ orange & 0.885 (\textcolor{forestgreen}{0.041}) & 0.874 (\textcolor{forestgreen}{0.005}) & 0.885 (\textcolor{forestgreen}{0.041}) & 0.032 (\textcolor{red}{0.032}) & 0.557 (\textcolor{forestgreen}{-0.433})  \\
      \midrule
      $\theta^B = 1$ & 0.879 (\textcolor{forestgreen}{0.027}) & 0.882 (\textcolor{forestgreen}{0.029}) & 0.879 (\textcolor{forestgreen}{0.027}) & 0.076 (\textcolor{forestgreen}{-0.062}) & 0.166 (\textcolor{red}{0.009})  \\
      $\;\;\;$ blue & 0.877 (\textcolor{forestgreen}{0.017}) & 0.876 (\textcolor{forestgreen}{0.008}) & 0.877 (\textcolor{forestgreen}{0.017}) & 0.392 (\textcolor{forestgreen}{-0.477}) & 0.073 (\textcolor{red}{0.073})  \\
      $\;\;\;$ orange & 0.882 (\textcolor{forestgreen}{0.038}) & 0.873 (\textcolor{forestgreen}{0.004}) & 0.882 (\textcolor{forestgreen}{0.038}) & 0.016 (\textcolor{red}{0.016}) & 0.666 (\textcolor{forestgreen}{-0.324})  \\
      \midrule
      $\theta^C = 1$ & 0.865 (\textcolor{forestgreen}{0.013}) & 0.873 (\textcolor{forestgreen}{0.020}) & 0.865 (\textcolor{forestgreen}{0.013}) & 0.208 (\textcolor{red}{0.070}) & 0.062 (\textcolor{forestgreen}{-0.095})  \\
      $\;\;\;$ blue & 0.854 (\textcolor{red}{-0.006}) & 0.868 (0.000) & 0.854 (\textcolor{red}{-0.006}) & 0.918 (\textcolor{red}{0.049}) & 0.001 (\textcolor{red}{0.001}) \\
      $\;\;\;$ orange & 0.877 (\textcolor{forestgreen}{0.033}) & 0.877 (\textcolor{forestgreen}{0.008}) & 0.877 (\textcolor{forestgreen}{0.033}) & 0.074 (\textcolor{red}{0.074}) & 0.389 (\textcolor{forestgreen}{-0.601}) \\
      \midrule
      $\theta^A = \theta^B = \theta^C$ & 0.883 (\textcolor{forestgreen}{0.031}) & 0.883 (\textcolor{forestgreen}{0.030}) & 0.883 (\textcolor{forestgreen}{0.031}) & 0.137 (\textcolor{forestgreen}{-0.001}) & 0.098 (\textcolor{forestgreen}{-0.059}) \\
      $\;\;\;$ blue & 0.884 (\textcolor{forestgreen}{0.024}) & 0.873 (\textcolor{forestgreen}{0.005}) & 0.884 (\textcolor{forestgreen}{0.024}) & 0.560 (\textcolor{forestgreen}{-0.309}) & 0.032 (\textcolor{red}{0.032}) \\
      $\;\;\;$ orange & 0.881 (\textcolor{forestgreen}{0.037}) & 0.875 (\textcolor{forestgreen}{0.006}) & 0.881 (\textcolor{forestgreen}{0.037}) & 0.057 (\textcolor{red}{0.057}) & 0.450 (\textcolor{forestgreen}{-0.540}) \\
\bottomrule
\end{tabular*}
\caption {\label{tab:experiments:results:result-table-synthetic} We report the performance metrics and error rates of the synthetic experiment after \methodname has been applied, and the corresponding relative improvements or deteriorations (green or red values in parenthesis, respectively).
	The first line always refers to the whole data set, whereas ``blue'' and ``orange'' contain results disaggregated by group.
	The top row shows metrics before \methodname has been applied.
}
\vspace{-8mm}
\end{table}
\endgroup

Table~\ref{tab:experiments:results:result-table-synthetic} shows how \methodname affects classification performance and fairness.
We understand fairness in terms of the three fairness objectives we seek to meet and therefore report metric and error rate differences in order to judge \methodname's performance.
The table reports the absolute values for each metric after \methodname has been applied, together with the changes relative to when \methodname is not applied (in parentheses).
The top row shows the metrics before \methodname has been applied.

Observe the results for our first experimental setting with $\theta^A = 1$.
We expect an overall performance increase, as the algorithm calibrates the predictions with respect to actual ground truth evaluation.
We also expect the effect to be more pronounced for the orange group, because of the larger error between predicted and ground truth scores for that group.
Both expectations are confirmed in the results.
The chance of the orange group to be labeled positive is now 5.7\% of the blue group, which marks an improvement of more than two orders of magnitude.

Next, observe the results for the second and third fairness objectives where $\theta^B=1$, and $\theta^C=1$, respectively (corresponding to fairness criteria B) and C), respectively).
When fulfilling criterion B), fair score distributions for ground truth negative individuals from the two groups should overlap (Figure \ref{fig:experiments:results:synthetic}b, top row).
This corresponds to a noteable FPR decrease for the blue group, which had a high base FPR including many false positive ground truth negative individuals scoring just slightly above the decision boundary whom become true negative after \methodname is applied.
This also corresponds to a slight FPR increase for the orange group, which had a base FPR of zero and ground truth negative individuals scoring far enough below the decision boundary that only a few were mapped by \methodname above the decision boundary to become false positives.
The results confirm our expectations.
For the blue group, \methodname achieves an FPR of 39.2\%, which marks an improvement w.r.t. the original FPR of 47.7\%.
We also see a slight increase of FPR for the orange group (1.6\%).
%
%This also explains why the $\text{FNR}=66.6\%$ for the orange group is still rather high: the original model labeled almost every orange individual as negative, thus yielding a $\text{FNR}=99\%$.
%
After \methodname is applied the FNR shows an improvement of 32.4\%.
Overall the probability of the orange group to receive a positive label is 5.3\% of the blue group for $\theta^B=1$.
When fulfilling criterion C), we expect the fair score distributions for true positive individuals to match (Fig.~\ref{fig:experiments:results:synthetic}c), thus the false negative rates should improve, particularly for the orange group since most blue individuals were predicted positive by the original model.
Again our expectations are confirmed by the results and declines in performance and error rates remain relatively small for both groups.
Overall, the probability of the orange group for a positive label is 7.7\% of the blue group.

\FloatBarrier
Last, observe the last experimental setting $\theta^A = \theta^B = \theta^C$, which corresponds to a compromise between the three mutually exclusive fairness criteria.
\noindent The results in Table~\ref{tab:experiments:results:result-table-synthetic} show that \methodname yields a compromise between calibration, balance for the true negatives, and balance for the true positives.
It achieves similar performance improvements as aiming for calibration only ($\theta^A=1$), but better FPR and FNR improvements.
Compared to the balance criteria, this setting achieves better performance improvements, but only slightly worse FPR and FNR.
Overall, the probability of the orange group to receive a positive label is 9.7\% of the blue group for $\theta^A=\theta^B=\theta^C$.

\subsection{Experiments on the \compas data set by~\cite{angwin2016machine}}
\subsubsection{Data set}
\compas (Correctional Offender Management Profiling for Alternative Sanctions) is a commercial tool developed by Northpointe, Inc.\ to assess a criminal defendant's likelihood of recidivating within a certain period of time.
Based on the promise to enhance fairness in judicial decision making, \compas is used in several US states as a decision aid for judges, \eg in parole cases.
In 2016, \citet{angwin2016machine} published an analysis of the tool based on a data set of criminal defendants from Broward County, Florida, in which they found the tool to be biased against certain groups (more on this below).

From the data---as it was published by~\cite{angwin2016machine}---we use \texttt{decile\_score} (integers $\{1, 2, ..., 10\}$) as predicted scores, and \texttt{two\_year\_recid} (boolean) as ground truth labels.
Additionally, we construct groups based on sex, race, and age category using the features \texttt{sex}, \texttt{race}, and \texttt{age\_cat}.
To increase race group sizes, we merge races `Native American' and `Asian' into `Other', leaving four race groups: `Caucasian,' `African-American,' `Hispanic,' and `Other.'
These three features, \ie the predicted score, the ground truth label, and the group, form the input for \methodname.

To assess the impact of \methodname on the predictive performance and the group error rates, we perform an analysis similar to that carried out by~\citet{angwin2016machine}.
We also translate the predicted score into a binary label of high and low risk corresponding to predicted score $\geq 5$ and $< 5$, respectively.
This binarization is applied to both the predicted scores from the \compas data set and the resulting fair scores produced by \methodname.
Additionally, we calculate the probability of the disadvantaged groups to be assigned a high risk label relative to the advantaged group while correcting for the seriousness of their crime, previous arrests, and future criminal behavior.

%\begin{figure}[b!]
%	\vspace{-3mm}
%	\centering
%	\subfloat
%	[Normalized \compas scores of non-recidivants, disaggregated by gender. Note that bars are not stacked.
%	\label{fig:experiments:result:compas:origDecilesInnocent}]
%	{\includegraphics[width=.49\textwidth]{figs/compas/gender/predScoresPerGroupWithCondition_groundTruthLabel=0_AsHistograms.png}}\hfill
%	\subfloat
%	[Normalized \compas scores of recidivants, disaggregated by gender. Note that bars are not stacked.
%	\label{fig:experiments:result:compas:origDecilesReoffenders}]
%	{\includegraphics[width=.49\textwidth]{figs/compas/gender/predScoresPerGroupWithCondition_groundTruthLabel=1_AsHistograms.png}}
%	\vspace{-3mm}
%	\caption{ (Best seen in color).	Decile score distributions of the \compas model, disaggregated by ground truth and gender.
%		%
%		Fig.~\ref{fig:experiments:result:compas:origDecilesInnocent} depicts the scores for male (blue) and female (orange) individuals who \emph{did not} reoffend, Fig.~\ref{fig:experiments:result:compas:origDecilesReoffenders} depicts the scores for those who \emph{did} reoffend.
%		%
%	}
%	\label{fig:experiments:results:compas-orig}
%\end{figure}

\begin{wrapfigure}{R}{.55\textwidth}
	\vspace{-3mm}
	\begin{minipage}{\linewidth}
		\compas accuracy disaggregated by decile score and:

		%\\
%		\\
		\centering
		(a) gender
		\includegraphics[width=\linewidth]{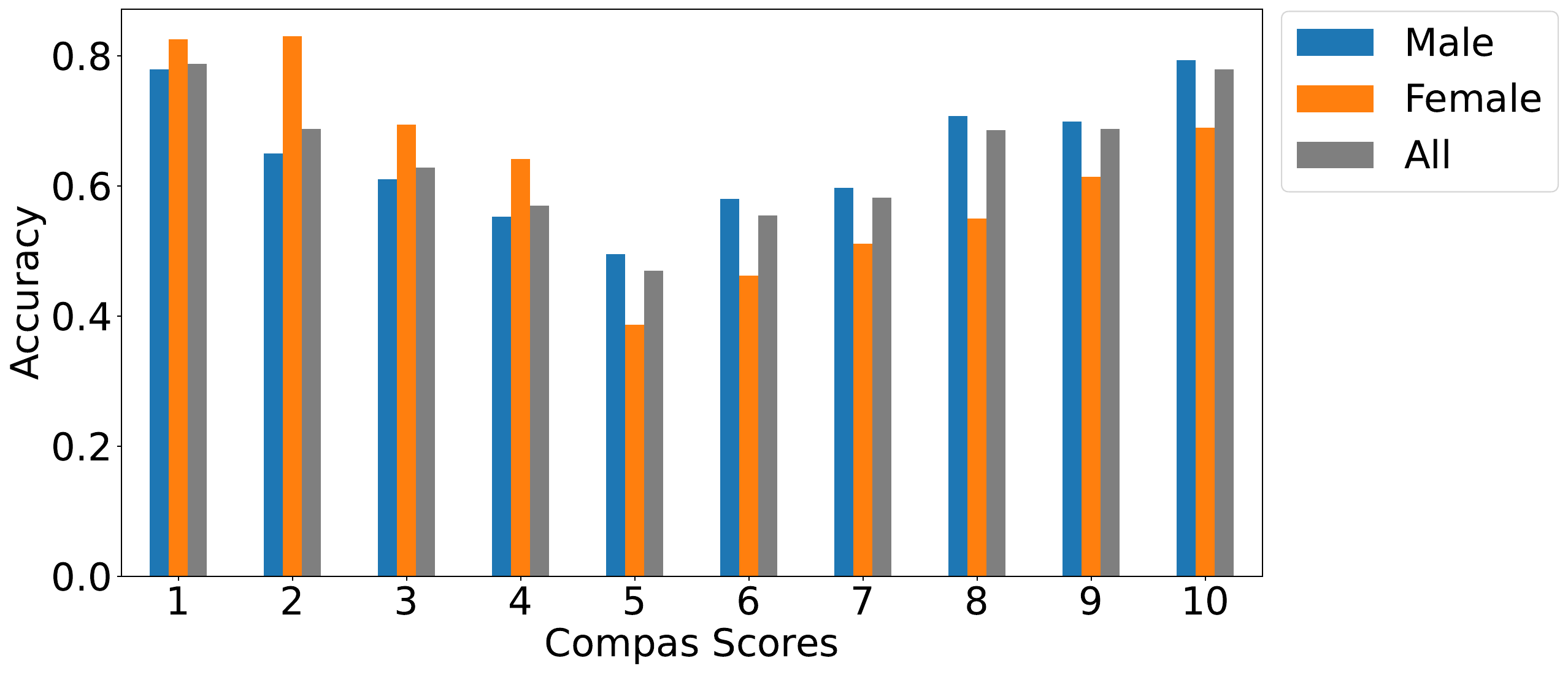}

		\centering
		(b) race
		\includegraphics[width=\linewidth]{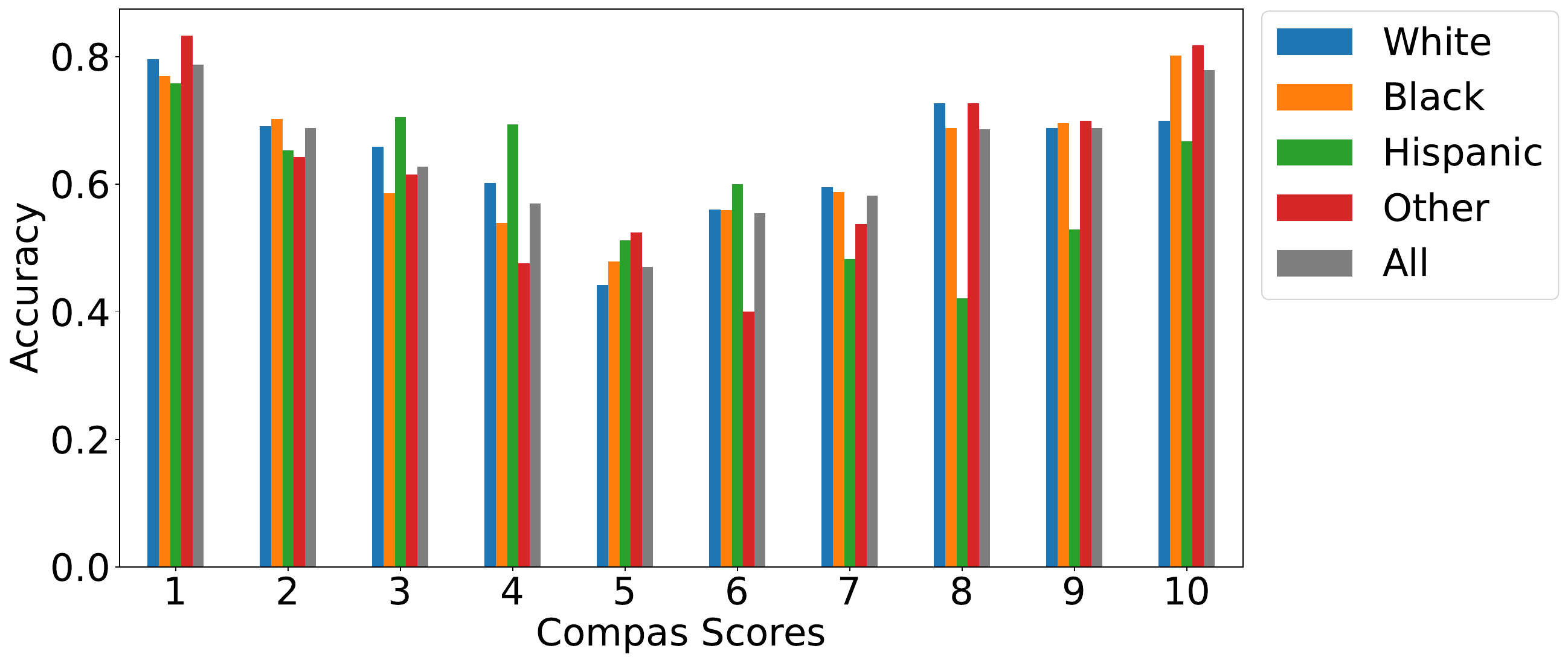}

		\centering
		(c) age
		\includegraphics[width=\linewidth]{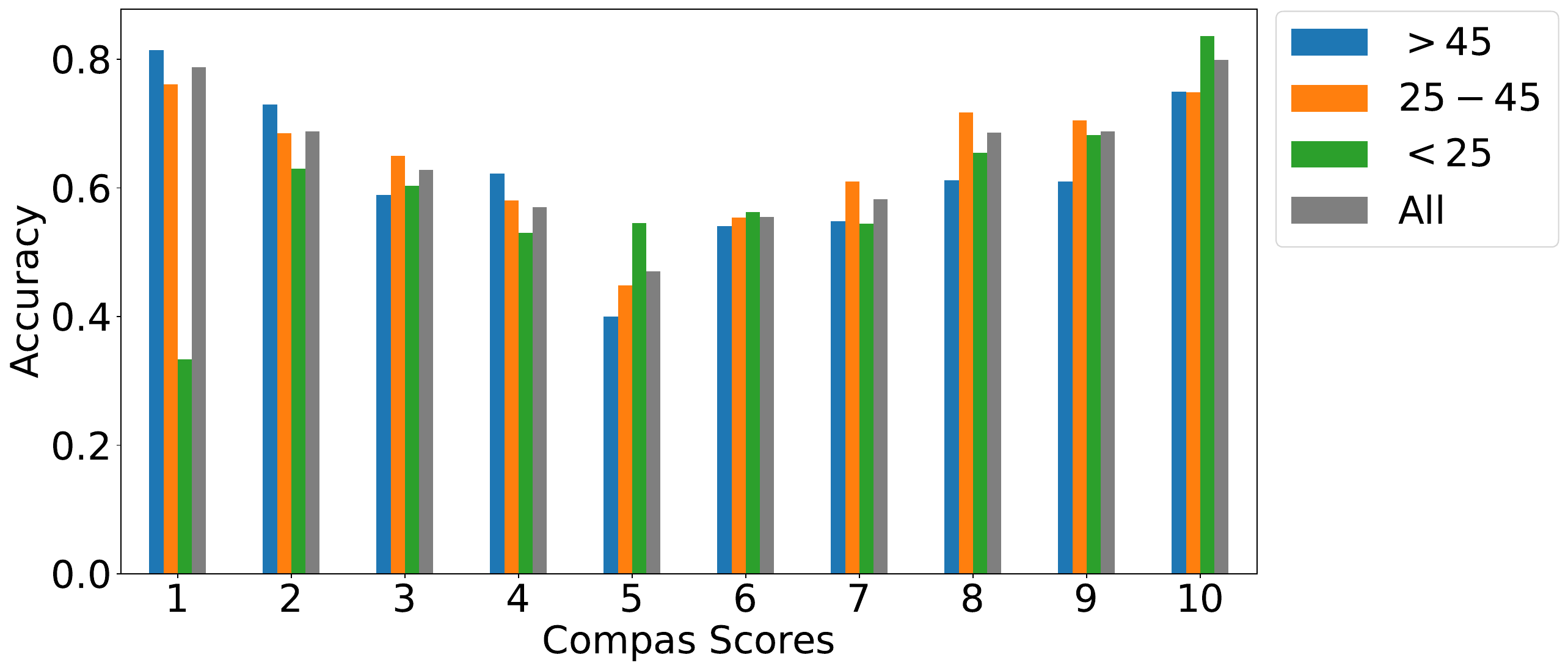}
	\end{minipage}
	\caption{(Best seen in color) Accuracy rates disaggregated by \compas score. Across all demographics, accuracy rates show that the mid-range scores 4--7 do not provide very meaningful insight on recidivism risk.
		%It is therefore questionable why these scores exist at all.
	}\label{fig:experiments:compas}
	\vspace{-1mm}
\end{wrapfigure}

\subsubsection{Performance Analysis of the Original Model}
\label{sec:experiments:compas:performance-orig}
To better interpret the results of applying \methodname to the \compas data set, we first provide a detailed performance analysis of the original \compas algorithm.
Fig.~\ref{fig:experiments:compas} shows the accuracy rates of the original model disaggregated by decile score and demographic group.
Grey bars mark the accuracy rates measured for the aggregated data set.
A striking insight is how poorly the model performs through the entire score range, but particularly for intermediate scores: within the range of 4--7, the accuracy rate for any demographic group hardly ever exceeds a level of 0.6.
One could argue, of course, that a triangular pattern of accuracy rates (high accuracy for edge scores 0 and 10, lower accuracy for scores near the decision boundary of 5) would be expected from a well-calibrated model.
After all, for a calibrated system, roughly 50\% of individuals receiving a \compas score of 5 (corresponding to a normalized score of 0.5) would reoffend.
Given that a \compas score $\geq 5$ corresponds to a prediction that an individual will reoffend, a calibrated system would achieve around 50\% accuracy for the 5th score decile.
It is important to understand, however, that if the model was well-calibrated we would expect to see accuracy rates of \emph{at least} 0.9 for scores 1 and 10, \emph{at least} 0.8 for scores 2 and 8, and so forth, requiring an accuracy rate of \emph{at least} 0.5 for score 0.5.
As shown, the model never achieves any of these minimum performance rates for any demographic group.

%Additionally, we see in Fig.~\ref{fig:experiments:results:compas-orig} further evidence that the \compas model is not calibrated:
%%
%For a well-calibrated model, one would expect $0-10\%$ of individuals who score in decile 1 to reoffend, and $90-100\%$ of individuals who score in decile 10 to reoffend.
%%
%As score increases, this would correspond to a linear decrease in non-recidivant frequency, and a linear increase in recidivant frequency.  \alex{actually, I guess the absolute frequency depends on the base rates and is not expected to change linearly (like we originally reasoned).  The part that should be linear with score is the proportion of recidivants vs all individuals in each score decile}
%%
%This is not the case both for ground-truth negatives and positives, but the model is particularly flawed for recidivants: scores from 0.1 -- 0.9 are more or less \emph{uniformly distributed}.
%%
%This means that, for recidivants, getting a score of 0.9 and getting a score of 0.1 \emph{is equally likely}, which would not be problematic if these probabilities were low.
%%
%Yet, receiving the highest score is only about twice as likely as getting any of the other scores.
%%
%Thus, recidivants face an overall probability of being labeled high-risk (scores 0.8--1.0) of about 36\%.
%%
%Ironically, their chance of being labeled low-risk (scores 0.1--0.4) is about the same.

Another interesting insight can be found by looking at the disparities of accuracy distributions across different demographic groups.
Observe Fig.~\ref{fig:experiments:compas}a: We see that women's accuracy is higher than men's in the lower scores, but vice versa for higher scores.
This means that women are predominantly misclassified when they are assigned high scores, while men are predominantly misclassified when they are assigned low scores.
In other words, women are treated too harshly by the algorithm, while men are treated too gently (thus confirming the finding of \citet{angwin2016machine}).
The same is true for Hispanics in Fig.~\ref{fig:experiments:compas}b, whose accuracy for the low scores is much higher than the one for the high score range.
Surprisingly, Fig.~\ref{fig:experiments:compas}c reveals that young people, even though already being assigned relatively high scores, still seem to be treated too gently by the algorithm.

\subsubsection{Experimental Results when Applying \methodname}\label{expCOMPAS}

Our experimental results using \methodname on the \compas data set mostly confirm an important hypothesis: our algorithm cannot fix an inherently flawed model (garbage in, garbage out).
As described earlier, we assessed accuracy, weighted average precision and weighted average recall, as well as false positive and false negative rates, total and disaggregated by demographics (similarly to results shown in Table~\ref{tab:experiments:results:result-table-synthetic}).
Because the nominal changes were negligible, we performed this same analysis disaggregated by \compas score (as in Fig.~\ref{fig:experiments:compas}).
Together with the analysis of the transport maps, this revealed the following: to fulfill fairness criteria B) and C), \methodname does not change the scores drastically.
This is understandable since \methodname does not correct for \emph{accurate} scores, but only for \emph{equal distributions}, as soon as $\theta^A=0$.
Thus, only a few individuals with original scores of 4 or 5 get reclassified from low to high risk, and vice versa.
Those with original scores 1--4, or 6--10, turn out not to cross the decision boundary between the low and high risk class and therefore, none of the metrics changes for these score ranges.

Therefore, we exemplarily report results disaggregated by gender and \emph{for score 5 only} in Table~\ref{tab:experiments:results:result-table-compas}, to showcase the behavior of \methodname on the \compas data set.
We see that the algorithm behaves as expected and does, in fact, improve performance and error rates the way we expect it to.
However, since the original model performance is so low (see first row section in Table~\ref{tab:experiments:results:result-table-compas}), we would recommend to rather abandon this model entirely.
It seems futile to change random guessing into fair random guessing.

\setlength{\tabcolsep}{1pt}
\begingroup
\begin{table}[t]
	\footnotesize
	\vspace{-3mm}
	\begin{tabular*}{\textwidth}{@{\extracolsep{\fill}}*{6}{llllll}}
		\toprule
		\begin{tabular}{@{}c@{}}\compas  \\ Data Set\end{tabular} & \multicolumn{3}{c}{Performance} & \multicolumn{2}{c}{Error Rates}   \\
		\cline{2-4} \cline{5-6}
		& Accur. ($\Delta$)  & \begin{tabular}{@{}c@{}}Weighted Avg. \\ Precision ($\Delta$)\end{tabular} & \begin{tabular}{@{}c@{}}Weighted Avg. \\ Recall ($\Delta$)\end{tabular}	& FPR ($\Delta$) & FNR ($\Delta$) \\    \midrule
		before \methodname & 0.470 & 0.751 & 0.470 & 1.000 & 0.000 \\
		$\;\;\;$ male & 0.495 & 0.750 & 0.495 & 1.000 & 0.000 \\
		$\;\;\;$ female & 0.387 & 0.763 & 0.387 & 1.000 & 0.000 \\
		\midrule
		$\theta^A = 1$ & 0.521 (\textcolor{forestgreen}{0.051}) & 0.751 (0.000) & 0.521 (\textcolor{forestgreen}{0.051}) & 0.739 (\textcolor{forestgreen}{-0.261}) & 0.185 (\textcolor{red}{0.185})  \\
		$\;\;\;$ male & 0.495 (0.000) & 0.750 (0.000) & 0.495 (0.000) & 1.000 (0.000) & 0.000 (0.000)  \\
		$\;\;\;$ female & 0.613 (\textcolor{forestgreen}{0.226}) & 0.763 (0.000) & 0.613 (\textcolor{forestgreen}{0.226}) & 0.000 (\textcolor{forestgreen}{-1.000}) & 1.000 (\textcolor{red}{1.000})  \\
		\midrule
		$\theta^B = 1$ & 0.530 (\textcolor{forestgreen}{0.060}) & 0.751 (0.000) & 0.530 (\textcolor{forestgreen}{0.060}) & 0.000 (\textcolor{forestgreen}{-1.000}) & 1.000 (\textcolor{red}{1.000})  \\
		$\;\;\;$ male & 0.505 (\textcolor{forestgreen}{0.010}) & 0.750 (0.000) & 0.505 (\textcolor{forestgreen}{0.010}) & 0.000 (\textcolor{forestgreen}{-1.000}) & 1.000 (\textcolor{red}{1.000})  \\
		$\;\;\;$ female & 0.613 (\textcolor{forestgreen}{0.226}) & 0.763 (0.000) & 0.613 (\textcolor{forestgreen}{0.226}) & 0.000 (\textcolor{forestgreen}{-1.000}) & 1.000 (\textcolor{red}{1.000})  \\
		\midrule
		$\theta^C = 1$ & 0.530 (\textcolor{forestgreen}{0.060}) & 0.751 (0.000) & 0.530 (\textcolor{forestgreen}{0.060}) & 0.000 (\textcolor{forestgreen}{-1.000}) & 1.000 (\textcolor{red}{1.000})  \\
		$\;\;\;$ male & 0.505 (\textcolor{forestgreen}{0.010}) & 0.750 (0.000) & 0.505 (\textcolor{forestgreen}{0.010}) & 0.000 (\textcolor{forestgreen}{-1.000}) & 1.000 (\textcolor{red}{1.000})  \\
		$\;\;\;$ female & 0.613 (\textcolor{forestgreen}{0.226}) & 0.763 (0.000) & 0.613 (\textcolor{forestgreen}{0.226}) & 0.000 (\textcolor{forestgreen}{-1.000}) & 1.000 (\textcolor{red}{1.000})  \\
		\midrule
		$\theta^A = \theta^B = \theta^C$ & 0.530 (\textcolor{forestgreen}{0.060}) & 0.751 (0.000) & 0.530 (\textcolor{forestgreen}{0.060}) & 0.000 (\textcolor{forestgreen}{-1.000}) & 1.000 (\textcolor{red}{1.000})  \\
		$\;\;\;$ male & 0.505 (\textcolor{forestgreen}{0.010}) & 0.750 (0.000) & 0.505 (\textcolor{forestgreen}{0.010}) & 0.000 (\textcolor{forestgreen}{-1.000}) & 1.000 (\textcolor{red}{1.000})  \\
		$\;\;\;$ female & 0.613 (\textcolor{forestgreen}{0.226}) & 0.763 (0.000) & 0.613 (\textcolor{forestgreen}{0.226}) & 0.000 (\textcolor{forestgreen}{-1.000}) & 1.000 (\textcolor{red}{1.000})  \\
		\bottomrule
	\end{tabular*}
	\caption {\label{tab:experiments:results:result-table-compas} Evaluation of \methodname when applied to the \compas data set, disaggregated by gender.
	These are results only for an original \compas score of 5, because only individuals close to the decision boundary turn out to get reclassified after a score reassignment through \methodname. Indeed, we find that for score ranges 1--4, and 6--10, neither error rates nor performance metrics change at all (see Subsection~\ref{expCOMPAS} for an explanation why this is expected).
	Note also that, for one particular score, everybody is classified either positive or negative and the per-group error rates are thus either 0 or 1.}
	\vspace{-6mm}
\end{table}
\endgroup

\subsection{E-commerce data set from Zalando}

\subsubsection{Data set}
For our last experiments, we use data from the e-commerce platform Zalando, one of Europe's largest fashion and lifestyle retailers, operating in 25 European countries.
We collected a data set of \np{81048} products \textcolor{maincolor}{for-sale} on 27th October 2021, which belong to four different women clothes categories with a similar price range: skirts, jeans, trousers, and knitwear.
Each data point contains the following information: a brand, a ranking score, \textcolor{maincolor}{number of impressions (how often the product was seen), and clicks.
Impressions and clicks correspond to the week following 27th October, 2021.}
Products are labeled as relevant (ground truth positive) if the ratio of clicks to impressions is above a certain threshold, meaning that they were of interest to users when seen by the customer.
All other products are labeled not relevant (ground truth negative).
Groups are assigned \textcolor{maincolor}{based on brand visibility calculated as the average number of impressions over all products per brand per category}.
When this average is below the median, a brand is said to belong to the \texttt{`low'} \textcolor{maincolor}{brand visibility group.
When this average is above the third quartile, a brand is said to belong to the \texttt{`high'} brand visibility group.}
Products of brands with visibility between the \texttt{low} and \texttt{high} group thresholds are discarded, leaving \np{62461} products.
70.5\% of them belong to \texttt{high} visibility brands.
Note that the ranking scores are produced by a learning-to-rank model trained daily to optimize a surrogate normalized Discounted Cumulative Gain (nDCG) loss~\cite{jarvelin2002ndcg}, but not to classify products according to the ground truth labeling we defined above.
However, because the nDCG relevance labels are based on customer interactions (\ie implicit feedback), it is reasonable to assume that a good ranking model should be able to identify which product is likely to be clicked.
Hence, the scores of an accurate click-through-rate classifier should provide a decent ranking.
We observe in our data that, irrespective of whether \texttt{high} visibility products are relevant or not, they receive higher ranking scores compared to \texttt{low} visibility products.
There are many reasons external to the ranking algorithm why customers would prefer well-known and highly visible brands: these brands have higher marketing budgets, they can afford a more aggressive pricing strategy thanks to economies of scale, and they benefit from the Matthew effect of accumulated advantage~\cite{perc2014matthew}.
Moreover, it is in the commercial interest of the platform to highlight such best-selling items.
At the same time, the platform may not want to reinforce the status quo but rather provide a level playing field for all brands to compete fairly.
A level playing field makes it easier to attract new brands on the platform, offers a more diverse assortment to a potentially larger customer base, and can be seen as a good step to address the legal requirements \textcolor{maincolor}{regarding competition law and the Digital Markets Act} discussed in \textcolor{maincolor}{section}~\ref{sec:legal:ecommerce}.

\subsubsection{Results}

\setlength{\tabcolsep}{1pt}
\begingroup
\begin{table}[b!]
	\footnotesize
	%\vspace{-5mm}
	\begin{tabular*}{\textwidth}{@{\extracolsep{\fill}}*{6}{llllll}}
	\toprule 
	\begin{tabular}{@{}c@{}}Zalando \\ Data Set\end{tabular} & \multicolumn{3}{c}{Performance} & \multicolumn{2}{c}{Error Rates}   \\
	\cline{2-4} \cline{5-6}
	& Accur. ($\Delta$)  & \begin{tabular}{@{}c@{}}Weighted Avg. \\ Precision ($\Delta$)\end{tabular} & \begin{tabular}{@{}c@{}}Weighted Avg. \\ Recall ($\Delta$)\end{tabular}	& FPR ($\Delta$) & FNR ($\Delta$) \\    \midrule
	before \methodname & 0.624 & 0.623 & 0.192 & 0.079 & 0.808 \\
	$\;\;\;$ high & 0.586 & 0.622 & 0.238 & 0.122 & 0.762 \\
	$\;\;\;$ low & 0.717 & 0.747 & 0.012 & 0.002 & 0.988 \\
	\midrule 
	$\theta^A = 1$ & 0.654 (\textcolor{forestgreen}{0.030}) &                                     0.585 (\textcolor{red}{-0.039}) &                                     0.518 (\textcolor{forestgreen}{0.327}) &                                     0.252 (\textcolor{red}{0.173}) &                                     0.482 (\textcolor{forestgreen}{-0.327})\\
	$\;\;\;$ high & 0.623 (\textcolor{forestgreen}{0.038}) &                                     0.585 (\textcolor{red}{-0.037}) &                                     0.604 (\textcolor{forestgreen}{0.366}) &                                     0.361 (\textcolor{red}{0.239}) &                                     0.396 (\textcolor{forestgreen}{-0.366})\\
	$\;\;\;$ low & 0.729 (\textcolor{forestgreen}{0.012}) &                                     0.576 (\textcolor{red}{-0.171}) &                                     0.186 (\textcolor{forestgreen}{0.174}) &                                     0.055 (\textcolor{red}{0.053}) &                                     0.814 (\textcolor{forestgreen}{-0.174})\\
	\midrule 
	$\theta^B = 1$ & 0.621 (\textcolor{red}{-0.003}) &                                     0.625 (\textcolor{forestgreen}{0.001}) &                                     0.169 (\textcolor{red}{-0.022}) &                                     0.070 (\textcolor{forestgreen}{-0.010}) &                                     0.831 (\textcolor{red}{0.022})\\
	$\;\;\;$ high & 0.577 (\textcolor{red}{-0.008}) &                                     0.625 (\textcolor{forestgreen}{0.003}) &                                     0.189 (\textcolor{red}{-0.050}) &                                     0.095 (\textcolor{forestgreen}{-0.027}) &                                     0.811 (\textcolor{red}{0.050})\\
	$\;\;\;$ low & 0.725 (\textcolor{forestgreen}{0.008}) &                                     0.623 (\textcolor{red}{-0.124}) &                                     0.095 (\textcolor{forestgreen}{0.083}) &                                     0.023 (\textcolor{red}{0.021}) &                                     0.905 (\textcolor{forestgreen}{-0.083})\\
	\midrule 
	$\theta^C = 1$ & 0.622 (\textcolor{red}{-0.003}) &                                     0.623 (\textcolor{red}{-0.001}) &                                     0.177 (\textcolor{red}{-0.015}) &                                     0.073 (\textcolor{forestgreen}{-0.006}) &                                     0.823 (\textcolor{red}{0.015})\\
	$\;\;\;$ high & 0.579 (\textcolor{red}{-0.007}) &                                     0.625 (\textcolor{forestgreen}{0.003}) &                                     0.196 (\textcolor{red}{-0.043}) &                                     0.099 (\textcolor{forestgreen}{-0.023}) &                                     0.804 (\textcolor{red}{0.043})\\
	$\;\;\;$ low & 0.725 (\textcolor{forestgreen}{0.008}) &                                     0.608 (\textcolor{red}{-0.139}) &                                     0.104 (\textcolor{forestgreen}{0.092}) &                                     0.027 (\textcolor{red}{0.025}) &                                     0.896 (\textcolor{forestgreen}{-0.092})\\
	\midrule 
	$\theta^A = \theta^B = \theta^C$ & 0.633 (\textcolor{forestgreen}{0.009}) &                                     0.617 (\textcolor{red}{-0.007}) &                                     0.259 (\textcolor{forestgreen}{0.067}) &                                     0.110 (\textcolor{red}{0.031}) &                                     0.741 (\textcolor{forestgreen}{-0.067})\\
	$\;\;\;$ high & 0.594 (\textcolor{forestgreen}{0.009}) &                                     0.618 (\textcolor{red}{-0.004}) &                                     0.294 (\textcolor{forestgreen}{0.055}) &                                     0.153 (\textcolor{red}{0.031}) &                                     0.706 (\textcolor{forestgreen}{-0.055})\\
	$\;\;\;$ low & 0.727 (\textcolor{forestgreen}{0.010}) &                                     0.601 (\textcolor{red}{-0.146}) &                                     0.124 (\textcolor{forestgreen}{0.112}) &                                     0.033 (\textcolor{red}{0.031}) &                                     0.876 (\textcolor{forestgreen}{-0.112})\\
	\bottomrule
	\end{tabular*}
	\caption {\label{tab:experiments:results:result-table-zalando} We report the performance metrics and error rates of the experiment on Zalando data after \methodname has been applied, and the corresponding relative improvements or deteriorations (green or red values in parenthesis, respectively).
		The first line always refers to the whole data set, whereas ``high'' and ``low'' contain results disaggregated by group.
		The top row shows metrics before \methodname has been applied.
	}
	\vspace{-8mm}
\end{table}
\endgroup

%\begin{figure}[b!]
	%\vspace{0mm}
	%\centering
	%\includegraphics[width=\textwidth]{figs/figure2-paper-edit.png}
	%\caption{{\color{blue}(Best seen in color). Prediction scores before and after applying \methodname to shift scores to achieve (a-c) each of the three fairness criteria, respectively, and (d) an equal mix of all three fairness criteria. The prediction scores before \methodname are the same as in Fig.~\ref{fig:experiments:data sets:synthetic} but scaled to values between 0 and 1. The score distributions (top row) have been smoothed using kernel density estimation to make comparison between distributions easier to interpret. The discrete score transport plots (bottom row) shows how mixing of fairness criteria (d) yields a score transport somewhere in between that corresponding to each respective fairness critera.}}
	%\vspace{-3mm}
	%\label{fig:experiments:results:synthetic}
%\end{figure}

\begin{figure}[t!]
	%\vspace{-10mm}
	\centering
	\includegraphics[width=0.8\textwidth]{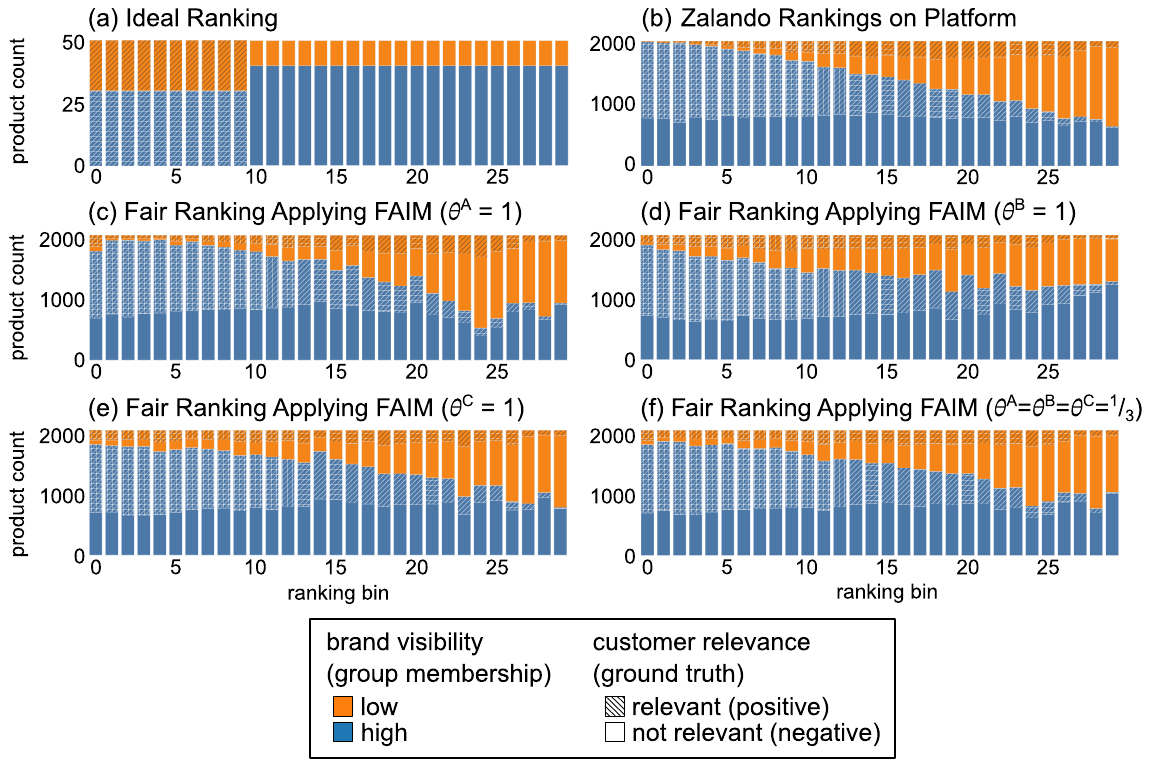}
	\vspace{-75mm}
	\caption{(Best seen in color).
		Product rankings with \texttt{low} and \texttt{high} visibility brands under different settings.
		\textcolor{maincolor}{Color indicates the average overall brand visibility of a product (group membership: low and high), whereas fill pattern indicates customer relevance (ground truth: relevant and not relevant).
		Fig.~\ref{fig:experiments:results:zalando}a exemplifies the outcome of an ideal ranker that ranks products by relevance irrespective of brand (and brand visibility).
		Fig.~\ref{fig:experiments:results:zalando}b shows the distribution of relevant vs not relevent products from \texttt{low} and \texttt{high} visibility brands as shown by Zalando's ranking algorithm by the time of data collection.
		We see that \texttt{high} visibility brands do indeed have an advantage over \texttt{low} visibility brands, as there are almost no products from the \texttt{low} brand visibility group in the first few bins.
		Fig.~\ref{fig:experiments:results:zalando}c--f show how the ranking changes after \methodname is applied with different values for $\theta^A$, $\theta^B$, and $\theta^C$.
		Depending on our preference for calibration, balance for the positive class, or balance for the negative class, we see varying degrees of ranking improvement for relevant products from the \texttt{low} brand visibility group.}
	}
	\vspace{-3mm}
	\label{fig:experiments:results:zalando}
\end{figure}

From Table~\ref{tab:experiments:results:result-table-zalando} we see that low visibility brands do indeed have a relative disadvantage over high visibility brands.
The low recall, and the high false negative rate indicate that many relevant products from the \texttt{low} group are not shown to the customer.
This situation is improved by \methodname in all four scenarios, albeit to varying degrees.
Our results also show that the algorithm yields expected improvements: When criterion A) is desired, accuracy and recall is improved for both groups.
Pursuing criterion B) or C) levels the respective error rates.
Finally, a compromise between all three criteria is found when setting $\theta^A = \theta^B = \theta^C$.

For a better understanding on the effect of \methodname on the rankings, observe Fig.~\ref{fig:experiments:results:zalando}.
In Fig.~\ref{fig:experiments:results:zalando}a we show a fictive example for an ideal ranking: imagine a data set with \textcolor{maincolor}{brand visibility groups.
The low brand visibility group has 400 products: 200 relevant and 200 non relevant. The high brand visibility group has 1100 products: 300 relevant and 800 non relevant.}
If we assume a \textcolor{maincolor}{ideal} ranker, we can sort those products simply by their relevance score.
\textcolor{maincolor}{When grouping the rankings in bins of \np{50} products, we would expect the top 10 bins (top 500 ranked products) to contain only relevant products, and the next 20 bins (next 1000 ranked products) to contain only non relevant products.}
The top 10 (relevant) bins should contain a high to low brand visibility ratio of 300/200, whereas the next 20 (non relevant) bins should contain a high to low brand visibility ratio of 800/200.
Fig.~\ref{fig:experiments:results:zalando}b shows the ranking as collected from the Zalando website on 27th October 2021, which, by virtue of being real, is not ideal.
Since each bin contains more than \np{2000} products, it is enough to focus on the first bins, as products from other bins are shown very rarely unless customers actively look for them.
We see that the first five bins contain mostly \texttt{high} visibility brands, meaning that relevant products from \texttt{low} visibility brands are rarely shown to the customer unless they start to use product filters or full text search.
Again, there are various reasons why a customer would prefer \textcolor{maincolor}{high visibility brands over low visibility brands}, but if we wanted to (or had to because of legislation) mitigate the visibility disparities, \methodname provides a convenient approach (see Section~\ref{sec:legal:ecommerce} for a discussion on \methodname's legal meaning for an e-commerce scenario such as ours).
Figures~\ref{fig:experiments:results:zalando}c--~\ref{fig:experiments:results:zalando}f show rankings after \methodname has been applied with different values for $\theta^A$, $\theta^B$, and $\theta^C$.
In all cases, \methodname distributes \textcolor{maincolor}{high product ranking scores more evenly across both brand visibility groups}, particularly for relevant products, which is the interesting case for an e-commerce platform such as Zalando.

\section{Application of the Algorithm in Legal Scenarios}

Given the \emph{contextual agility} \cite{wachter2021fairness} of EU regulations with respect to the definition of fairness, it is advantageous to introduce a similar flexibility in the operationalization of fairness concepts. \methodname has a wide range of applications in scenarios in which the law compels the decision maker to make decisions in both an accurate \emph{and} a non-discriminatory way.

Importantly, this is of key importance for compliance with the newly enacted EU AI Act.  According to Article 9 AI Act, providers of high-risk AI systems need to assess and mitigate risks to health, safety and fundamental rights, such as non-discrimination. Similarly, according to Article 55, developers of certain particularly powerful foundation models need to do the same. Meanwhile, Article 15 prescribes sufficient accuracy and performance of high-risk AI systems. Law enforcement, credit scoring, hiring and medical AI qualify as high-risk AI applications, making compliance with Articles 9, 15 and 55 mandatory for any provider seeking to offer its model in the EU. However, as we shall discuss in detail in the following sections, mitigating risk across different fundamental rights is not always straightforward or unidimensional. For example, in some scenarios, reducing biased outcomes may lower performance. This complexity is not addressed in the AI Act itself, and underappreciated by current literature on the AI Act. Significantly, machine learning and compliance systems need to be able to navigate these subtleties to deploy AI in a legal way.

Technically speaking, this may result in a trade-off between competing fairness and accuracy measures embodying the contemplated criteria, \ie calibration; balance for the negative class; balance for the positive class.
Importantly, the weighting of the different measures via the $\theta$ interpolation parameters is a deeply normative choice which reflects competing visions of what is supposed to be considered fair, and legal, in a certain situation.
Depending on the domain of application, the achievement of calibration, the balancing of false positives or of false negatives might be considered most important.
\methodname affords the advantage of allowing the decision maker to flexibly adapt to these different scenarios by choosing different $\theta$ weights.

The broad discussion of different fairness metrics both in computer science and the law \cite{wachter2021bias,hellmann2020measuring,binns2018lessons,friedler2021impossibility,kim2017auditing,pessach2020fairness,mitchell2021,meel2021fairness} suggests that no single metric such as statistical parity, error balance or equalized odds will fit all contexts. This also applies to the fairness measures under discussion here \cite{corbett2018measure}.
As pointed out before, an important prerequisite for those metrics is the reliability of ground truth~\cite{bao2021compaslicated}.
If these data points are collected in a way that systematically disfavors one protective group, conditioning on ground truth data risks perpetuating imbalances included in them \cite{bao2021compaslicated,wachter2021bias,chouldechova2017fair}.
Within these constraints, ground truth measures like the ones underlying \methodname can nevertheless be helpful tools in many situations to capture normative desiderata, particularly if used alongside strategies to improve the correctness and representativeness of ground truth.

Generally speaking, choosing, for example, between a stronger balance for the negative or the positive class will depend on the respective consequences of misclassification (false negative and false positive predictions).
As a result of prediction errors, individual and social costs arise, and legal norms may be violated.
These costs and norms will differ widely depending on the deployment context. 

However, the AI Act, and other laws, do not explicitly conceptualize the trade-offs between different risks, leaving this complex balancing act primarily to the discretion of developers. Current legal literature is equally silent on this problem. This lack of guidance can create significant challenges for developers, who must navigate these trade-offs without a clear framework. Our results provide a structured approach for transparently addressing these trade-offs, helping developers comply with the AI Act while ensuring a balanced consideration of all relevant risks. This framework aims to support developers in making informed decisions that align with the objectives and compliance with the provisions of the AI Act, promoting fairness and effectiveness in AI system deployment.
In the following, we review three examples in which different normative considerations may lead a decision maker to adopt varying weights for the respective accuracy and non-discrimination measures: recidivism prediction and credit scoring -- both high-risk applications under the AI Act; and fair ranking according to the Digital Markets Acts (DMA).

\subsection{Recidivism Prediction}
In this paper, we consider recidivism prediction for technical and argumentative reasons: the original papers by \citet{kleinberg2016inherent} and \citet{chouldechova2017fair}showing incompatibility between fairness metrics did use the COMPAS case, and we directly build on their work. Hence, we use COMPAS as an example to clarify how \methodname works. However, when considering recidivism prediction, the first questions that arise from a juridical perspective are whether one legally may, or policy-wise should, use algorithmic tools such as the COMPAS model to assess recidivism risk in Criminal Justice proceedings at all (see also \ref{sec:discussion:compas and missing data}). We do not wish to make any claim as to the legitimacy of the usage of these tools in this paper; there are, if anything, good reasons to be quite skeptical about it.

Under the EU AI Act, such systems qualify as high-risk systems (Annex III, points 6 and 8 AI Act); to the extent that they rely only on profiling or personality assessments, they are even prohibited (Article 5(1)(d) AI Act). High-risk systems need to be designed such as to simultaneously fulfill a criteria of (public) safety, personal freedom of individuals, fairness/non-discrimination (Article 9 AI Act) and performance criteria (Article 15 AI Act), triggering inherent legal and technical tensions.

As mentioned, COMPAS is a proprietary model used in pretrial settings in the US to determine whether potential offenders should be detained until their trial, based on a prediction of recidivism likelihood.
Similar models are currently employed in Canada, the UK, and Spain \cite[p. 122]{watch2019automating}.
While the use of such instruments has sparked significant controversy, and critique, in the legal literature \cite{starr2014rationalization,mayson2019bias,selbst2017policing,katyal2019accountability,eaglin2017recidivism,huq2019equity,pruss2021distortion}, courts such as the Wisconsin Supreme Court have condoned their use under certain safeguards,\footnote{ State v. Loomis, 881 N.W.2d 749 (Wis. 2016).} and their deployment is currently on the rise \cite{hamilton2021evaluating}; \cite[p. 122]{watch2019automating}.

Within the scope of this paper, we cannot offer an in-depth discussion of the promises and perils of algorithmic recidivism risk assessment (see, e.g., \cite{solow2019institutional,kleinberg2018human}, and \ref{sec:discussion:compas and missing data}).
Rather, we would like to point out that, to the extent that such tools are used at all, they must clearly fulfill minimum requirements seeking to safeguard normative and legal principles of the jurisdictions in which they are deployed.
Importantly, the impossibility theorems concerning various fairness metrics mentioned above apply equally in the case of algorithmic and human risk assessments.
While the influence of fairness considerations on human risk predictions, for example by a judge, remains difficult to elucidate, \methodname allows to make the relevant trade-offs transparent and to rank the involved fairness metrics in varying degrees of priority.

In training the COMPAS model, the developers chose to prioritize calibration, which inevitably led to differing false positive and false negative rates given imperfect prediction and differing base rates between the involved ethnic groups \cite{dieterich2016compas,larson2016we}.
While accuracy remains significant in recidivism risk assessment---and is actually quite low with the COMPAS algorithm \cite{dressel2018accuracy}---the core normative trade-off arguably occurs between the respective importance of equalizing false positive and false negative predictions.
On the one hand, one could argue that the prevention of false positive outcomes should be prioritized because, under the rule of law, it is of utmost importance not to detain any person without sufficient reason.
Under this reading, the greatest weight should be put on aligning false positive rates between groups, so that the burden of being unduly sent to prison is shared equally between the respective groups.
On the other hand, one could claim that matching false negative predictions is crucial because they unduly spare individuals time in prison, affording a significant individual advantage to them. Under this reading, that undeserved benefit should not accrue to any protected group to a greater extent.\footnote{ Note that the fact false negative predictions constitute a particular risk to individuals and to society at large, with potentially dangerous offenders roaming free, is likely irrelevant here: it should not matter to victims, nor to society, by members of what protected group re-offences are committed. Criterion B) balances the negative classes, but does not (directly) reduce their size.}

\methodname allows for the establishment of an intermediary position such that both balances, for the positive and the negative class, are fulfilled to an equal degree, even though only partially. To the extent that such instruments are used at all, and that the data they are based on are considered adequate, \methodname may therefore operationalize a policy compromise between factually irreconcilable goals.

From a different perspective, however, balancing false positives may be considered more important: losses loom larger than gains. Time in prison constitutes a highly significant restriction of personal freedom, interrupting private lives and careers, potentially endangering physical and mental health. Hence, lawmakers or judges may come to the conclusion that it is more important to share the burden of false positive predictions equally between groups than the unwarranted benefit of false negative predictions. \methodname then allows to prioritize the former criterion.

\subsection{Credit Scoring}
Entirely different normative considerations are present in the case of credit scoring. Under the EU AI Act, AI-based credit scoring is considered a high-risk activity, just like law enforcement. Again, in this domain, the risk assessment and mitigation framework in the AI Act, particularly Articles 9 and 55, introduces inherent tensions in balancing various risks. Providers of high-risk AI systems and of particularly powerful foundation models are required to mitigate risks to health, safety, and fundamental rights. According to Article 15, they also have to ensure sufficient accuracy and performance. However, addressing one type of risk may inadvertently exacerbate another. For instance, reducing discrimination or enhancing transparency in AI models can sometimes lower their performance. This performance dip can impact the attainment of health objectives in medical AI or default prevention in credit scoring, which are also crucial for supporting fundamental rights. Moreover, other types of legislation provide for even more legal constraints on credit scoring with a view to performance and non-discrimination, as discussed below.

Imagine a bank uses an ML-based scoring algorithm to assess the creditworthiness of loan applicants, as in our synthetic experiment. Increasingly, ML is indeed used for these purposes \cite{langenbucher2020responsible,langenbucher2021responsible}.
Here, accuracy facilitates so-called responsible lending, i.e., loan decisions in which the credit institution intends to ensure that the borrower is not overburdened by the repayment obligations.
After the financial crisis of 2008/09, responsible lending has become a cornerstone of financial law.
In the EU, for example, the specific obligation to lend responsibly is enshrined in Article 8 of the Consumer Credit Directive 2008/48/EC, and other EU law instruments install a comprehensive compliance and supervision regime demanding regular audits to ensure the accuracy of credit scoring models, in addition to the AI Act (Art. 174 et seqq. of the Capital Requirements Regulation 575/2013, CRR).
Hence, accuracy should certainly receive significant weight in the case of credit scoring. As Art. 174(1) CRR puts it, statistical models used by banks need to have ‘good predictive power’. This is equally demanded by Article 15 AI Act.

However, with default base rates usually differing between protected groups, high degrees of calibration will lead to an imbalance in the false positive or false negative rates between the groups. A positive label, in credit scoring, means that the loan request is denied, a negative label that it is granted (because the risk of default is low enough).
Clearly, false negative predictions may inflict financial damage on the lender if the credit cannot be repaid, but also potentially on the borrower, who may face financial penalties, eviction and future encumbrance due to a negative credit record.
False positive predictions, on the other hand, give rise to opportunity costs: the lender does not earn interest payments, and the borrower does not obtain access to credit.
Disparities concerning false negative or false positive rates affect access to credit or default rates between protected groups, and are therefore relevant for compliance with non-discrimination legislation.

In the EU, as mentioned, such risks must be assessed and mitigated by developers of high-risk AI systems under Art. 9 AI Act, such as in credit scoring. Furthermore, specific rules on indirect discrimination forbid even seemingly neutral practices putting protected groups at a particular disadvantage, unless these differences can be justified.\footnote{ See, e.g., Art. 2(b) of Directive 2004/113/EC; Art. 2(2)(b) of Directive 2000/43/EC.}
Higher false positive rates clearly constitute a disadvantage for the affected group; even higher false negative rates can be viewed as a burden on the more often wrongly classified group, however, as they entail the concrete risk of default.
Since accuracy and error rates cannot be balanced between protected groups at the same time under non-trivial conditions, the law will have to accept reasonable trade-offs between these competing and mutually exclusive obligations. It cannot and does not demand what is impossible (nemo ultra posse obligatur).
From the perspective of legal doctrine, this will arguably take the route of the justification of a possible disadvantage.
The damage inflicted must then be proportionate, given the reasons which can be advanced for prioritizing the other fairness criteria.

Given these preconditions, equalizing false negative credit predictions might be considered more important in the case of high-stakes loans (e.g., large sums and little collateral).
Here, a default would be particularly disruptive, and hence that burden should be shared equally between protected groups.
As a consequence, the $\theta$ weight of the balance for the positive class (Criterion C) should be higher than that for the negative class (Criterion B).

Conversely, in the case of low-stakes loans (e.g., low credit volume; sufficient collateral besides basic necessities of the borrower, e.g. beyond the house a family lives in; or limited personal liability in case of default), false positive predictions might be more deleterious than false negative ones because the damage is limited if the borrower defaults on the credit.
Wrongfully denying access to credit may then have larger opportunity costs.
Under such circumstances, the $\theta$ weight of the balance for the negative class (Criterion B) should be higher than that for the positive class (Criterion C).
As an example, calibration could be set to 0.5, balance for the negative class to 0.35, and balance for the positive class to 0.15. Legally speaking, the fact that an unequal rate of false negative predictions persists will then arguably be justified by the overriding importance of aligning false positive predictions and calibration between groups. In our view, this would also be a justifiable risk mitigation and compliance strategy under Article 9 AI Act.

\subsection{Fairness in E-Commerce Rankings}
\label{sec:legal:ecommerce}
Finally, our method may be used to implement emerging legal notions of fairness in e-commerce rankings. In this context, the AI Act does not provide any guidelines. However, particularly in EU law, a growing number of other provisions aim to safeguard the impartiality of rankings in online contexts. At the most general level, these rules have shifted from the prohibition of self-preferencing via a focus on transparency to, so far, under-researched concepts of fairness in the Digital Markets Act (DMA). To the best of our knowledge, our contribution constitutes the first attempt to operationalize the fair ranking provisions of the DMA on both a technical and a legal basis.

\subsubsection{Competition Law}
The oldest and best-known rule of fairness in e-commerce rankings is derived from the prohibition, in general competition law, to abuse a dominant position (Art. 102 TFEU in EU law). It has long been argued that many dominant online undertakings, such as Google or Amazon, need to be scrutinized under this provision due to their dual role as providers of online marketplaces – on which third parties directly sell their goods to consumers – and as direct sellers of goods on their platforms~\cite{padilla2020self,graef2019differentiated}. Hence, such platforms may be considered competitors of the very business customers they serve on their marketplaces (vertical integration). As a consequence, dominant online platforms must not unduly preference their own offers vis-à-vis those of their business customers in rankings they display as a result of consumer search queries. This rule has led to some of the most spectacular fines in recent EU competition law. For example, the General Court of the European Union, in November 2021, affirmed the 2017 decision by the EU Commission to fine Google with €2.4 billion for engaging in self-preferencing in its online comparison shopping service.\footnote{GCEU, Case T-612/17 (Google Shopping).} A similar issue is at stake in the Amazon buy box case, in which the Italian Competition Authority, in December 2021, imposed a record fine of €1.1 billion on several Amazon companies for tying access to the buy box to the use of Amazon's own logistics channel (Fulfillment by Amazon).\footnote{Italian Competition Authority, Case A528, Amazon, Press release, 9 December 2021.} 

The upshot of these rulings is that dominant platforms which serve a dual role of marketplace and seller must abstain from systematically tweaking their rankings to their own benefit. This rule against self-preferencing, deduced from Art. 102 TFEU and pertinent case law, may be considered a first substantive fairness element for the order of the ranking itself~\cite{podszun2021digital}. However, it only applies to undertakings which dominate a certain market.

\subsubsection{Transparency}
Often, though, it is difficult for outsiders, even for the business customers, to determine how these rankings come about. Two new provisions therefore install explicit transparency provisions for rankings in EU law \cite{hacker2022varieties,eifert2021taming}. First, Art. 5 of the so-called P2B (Platform to Business) Regulation (EU) 2019/1150 obliges online intermediaries and search engines, independent of their market power, to disclose the main parameters of ranking and their relative importance. The provision has been in effect since July 2020. It is supposed to foster the predictability and understanding of the ranking for business users, and to foster competition between different intermediaries with respect to the ranking parameters (Recital 24 of the P2B Regulation). In a similar fashion, second, the new Art. 6a of the Consumer Rights Directive (CRD) has required online marketplaces from the end of May 2022 on to divulge to consumers the main parameters for rankings based on consumer search queries as well as their relative importance. According to the new Art. 2(1)(n) of the Unfair Commercial Practices Directive, ‘online marketplace’ denotes any software operated by or on behalf of a trader which allows consumers to conclude distance contracts with other traders or consumers. Therefore, the rule does not apply to companies only selling goods directly to consumers, but is again independent of market power. 

Essentially, scholars have argued, these provisions introduce an obligation for the global explanation of the ranking model \cite{hacker2022varieties,grochowski2021algorithmic}. It may be fulfilled, for example, by using interpretable machine learning models (e.g., linear or logistic regression~\cite{rudin2019stop}) or by using global post-hoc explanations for black box models (e.g., DNNs) such as an average over local SHAP values~\cite{lundberg2017unified} or SpRAy~\cite{lapuschkin2019unmasking}. However, the provisions do not add any specific, substantive fairness criteria concerning the order of the ranking itself.

\subsubsection{The EU Digital Markets Act}
\label{sec:DMA}
The DMA, which has come into effect in the EU in 2023, takes these existing provisions yet one step further by introducing a novel fairness condition for rankings. Its importance is hard to overstate: failure to comply with the DMA provisions may be sanctioned by fines of up to 10\% of the global annual turnover of the offender (Art. 30(1) DMA), and up to 20\% in the case of a repeated violation (Art. 30(2) DMA). The DMA includes a whole range of provisions aimed at strengthening contestable markets and fair competition in online environments by designing specific rules applicable to so-called gatekeepers, in essence large online platforms.

\paragraph{Fairness in the DMA}
Art. 6(5) DMA specifically tackles rankings,\footnote{ In the final version, it applies, beyond rankings, also to indexing and crawling.} including when delivered via a virtual assistant (Recital 51 DMA; for example, a personal voice assistant). Art. 6(5) DMA first reiterates the prohibition of self-preferencing known from general antitrust law.\footnote{ For differences to Art. 102 TFEU, see, e.g.,~\cite{eifert2021taming}.}

The second sentence of Art. 6(5) DMA starts by specifying that rankings need to be transparent. As Recital 52 suggests, this could refer to the transparency provisions of the P2B Regulation. Importantly, Art. 6(5) DMA goes on to demand that gatekeepers must, in general, apply ‘fair and non-discriminatory conditions’ to their rankings. In the legal space, this has sparked a discussion on how these criteria may be interpreted~\cite{brouwer2021towards,hacker2022ki}.\footnote{For a discussion of non-discriminatory rankings in general competition law, see \cite{graef2019differentiated}.} So far, however, no clear guidance or consensus has emerged on what these additional fairness requirements for rankings under the DMA could mean for gatekeepers.

This raises the question of how the obligation to apply fair and non-discriminatory conditions to \emph{rankings} could be interpreted. Much remains uncertain in this area \cite{hacker2023regulating,eifert2021taming}. Courts might, however, conclude in the future that fairness and non-discrimination under the DMA clause is linked to accuracy, false positives and false negatives. In this case, we can smoothly operationalize the DMA fairness condition with \methodname. This is significant insofar as previous discussions of fairness in algorithmic contexts have invariably focused on non-discrimination law~\cite{zehlike2017fair,asudeh2019designing,singh2018fairness,zehlike2022fairness}, and not on fairness conditions in e-commerce regulation, such as the DMA.\footnote{The DMA applies to a range of digital services, so-called core platform services (Art. 2(2) DMA). These include social media and video-sharing services, for example; online intermediation, typical for e-commerce, is another core platform service according to Art. 2(2)(a) DMA.}

\paragraph{DMA Ranking with \methodname }  
\methodname allows to specifically and flexibly re-order gatekeeper rankings such that certain conditions are prioritized, depending on the exact, and as of yet unknown, binding interpretation of Art. 6(5) DMA. In the case of e-commerce rankings, one possible interpretation by courts could be that accuracy as well as an equal false positive rate are the most important parameters among the three conditions. Under such a reading, false negative predictions might be considered less important than false positive predictions as most consumers will not even see or engage with those items which are ranked at low positions. If a platform, preemptively or as a result of a court or agency order, wishes to avoid that criteria such as incumbent status play a role in the top-ranked products, $\theta$ values can be set accordingly. They could, for example, convey high priority to equalizing the false positive rate (Criterion B) between high-visibility and low-visibility brands.\footnote{To be sure, there may be valid reasons for relegating certain brands to low-visibility status, for example reasons relating to product quality. Nonetheless, one may wish to equalize error rates between the groups. For instance, platforms may seek to limit false negative rates in the low group since even such brands may, eventually, produce some high-quality items; see also next para.}

A second possible interpretation would focus on false negatives. False negative predictions unduly demote products in the ranking and thus deprive affected items of visibility and hence of transactions. If a regulatory agency or court found that the false-negative rate needs to be balanced between certain groups, \methodname can again be tuned to implement such a requirement by choosing corresponding $\theta$ values (high values for Criterion C).

In future work, we will seek to identify further distinctions which may be problematic under the DMA fairness clause and which may be remedied by \methodname. Importantly, the fairness clause in the DMA provides an opportunity to reflect upon fairness criteria in e-commerce rankings beyond general non-discrimination law and the prohibition of self-preferencing in competition law.

\section{Discussion and Limitations}
\textcolor{maincolor}{\subsection{Choosing Fairness Weights}}
\textcolor{maincolor}{In the three discussed use cases, decision makers may place differing emphasis on various fairness criteria depending on which undesirable error type is most relevant. In recidivism prediction, balancing both error types might appear important, but judges or lawmakers may weight false positive predictions more heavily if unwarranted detentions are considered a particularly severe intrusion into personal freedom. In credit scoring, decision makers may choose to weight false negative rates more highly in high-stakes loan scenarios, while low-stakes loans may justify placing greater emphasis on false positives to avoid unnecessarily denying credit access. In e-commerce rankings, decision makers may prefer to focus on equalizing false positives if both accuracy and fairness in the top-ranked products are critical, or choose to emphasize false negatives if it is necessary to ensure that products are not unfairly pushed to low-visibility positions.}

\textcolor{maincolor}{More generally, the selection of weights for fairness criteria in different practical applications will vary, as the real-world consequences of imbalance differ significantly across contexts. Decision makers need to consider not only what each error type entails in theory, but also what it means in the given environment.}

\textcolor{maincolor}{It is often useful to start with a baseline weighting of 1/3 each and then adjust these weights based on an analysis of the stakeholders’ priorities, the severity of potential harms, and the anticipated legal and ethical constraints. In settings where reputation management is critical, decision makers may give equal weighting to all three criteria by setting them to 0.333 each, ensuring that no single measure dominates. In scenarios where a specific error type may lead to legal liability, that criterion could receive a higher weight such as 0.6 or 0.7, with the remaining 0.2 or 0.3 split evenly between the other two criteria. In the following paragraphs, we discuss briefly considerations for hiring decisions, college admissions, and medical recommendations.}

\textcolor{maincolor}{Hiring decisions: In this context, false negatives (qualified candidates not hired) and false positives (unqualified candidates hired) can both have significant consequences. However, the harm of unfairly denying opportunities likely outweighs the cost of some suboptimal hires, unless the position is for a unique leadership role. Therefore, we would suggest generally emphasizing equal false negative rates between demographic groups (e.g. by sex, race) with a weight of 0.5 for Criterion C, 0.3 for equalizing false positives (Criterion B), and 0.2 for the Criterion A (calibration). This allows a company to draw on a more diverse set of candidates and experiment with different protected groups whose potential might otherwise be overlooked. If, however, the role is of particular importance for the company and quite unique (e.g., C-level position), avoiding false positives seems particularly desirable and justified. Here, Criterion B could therefore be pushed to 0.6 or 0.7, with the remaining weights split between the other two.}

\textcolor{maincolor}{College admissions: Similar to the general hiring scenario, the impact of a false negative (unfairly denying a student admission) is arguably greater than a false positive (admitting a less-qualified student). Students have a limited number of opportunities to attend top colleges. Conversely, admitting a few suboptimally qualified students (false positives) does not hurt colleges or universities much. Hence, equalizing false negative rates should be the priority, perhaps with a 0.6 weight on Criterion C, 0.2 on Criterion B, and 0.2 on overall calibration.}

\textcolor{maincolor}{Medical diagnosis and treatment recommendations: When applying a fairness algorithm like FAIM to medical diagnosis and treatment recommendations, setting the appropriate weights for the different fairness criteria is of utmost importance given the high-stakes nature of healthcare decisions \cite{mittelstadt2023unfairness}. False negative and false positive predictions can both lead to serious consequences for patients, so carefully considering the potential impact of each type of error is essential.
False negatives, where a patient with a serious condition is incorrectly given a clean bill of health, can result in delayed diagnosis and treatment. This is particularly concerning for conditions like cancer, where early detection and intervention are critical. If an algorithm has a higher false negative rate for certain demographic groups, such as older patients, it could lead to a disproportionate number of those individuals facing worse prognoses or even preventable deaths due to missed or delayed diagnoses.
False positives, on the other hand, can cause undue psychological stress, unnecessary medical procedures, and increased healthcare costs. Being incorrectly told you may have a serious illness can take a significant emotional and mental toll, even if subsequent tests reveal the initial diagnosis was wrong. False positives may also lead to invasive procedures like biopsies or surgeries that carry physical risks and financial burdens for the patient.}

\textcolor{maincolor}{The specific FAIM weights will depend on the characteristics of the medical condition in question and the relative risks associated with false negatives and false positives. For a life-threatening illness where a missed diagnosis could be deadly, a higher weight for Criterion C may be justified, such as 0.6, while Criterion B might be assigned a weight of 0.3, and overall calibration 0.1. For less severe conditions, where false positives pose a bigger problem, the weights could be adjusted accordingly.} 

\textcolor{maincolor}{Overall, aligning weight distributions with the underlying normative trade-offs relevant to the specific domain allows decision makers to create a more justified and context-sensitive approach to implementing fair classification schemes.}

\subsection{Normative Choices and Applicability}
To explicitly name the normative choices \methodname implies, we use the concepts by ~\citet{friedler2021impossibility} of two different metric spaces---\emph{construct space} and \emph{observable space}. Construct space denotes an idealized reality in which all feature values are correctly registered, and a decision is based on those features. However, decision makers usually do not have access to this realm of an objective truth. Rather, they need to base their model, and decisions, on observable space, which contains the feature values the decision maker has access to via measurement. Importantly, ground truth may correctly reflect, or at least closely approximate, construct space---or it may not. 

This implies a first limitation for our method. \methodname operates in observable space, and does not undo any harm that might arise from a potentially biased mapping between construct and observable space, e.g., by systematic measurement errors.
This is because we obtain our ground truth labels from observing behavior of scored individuals, such as: ``did someone who was labeled creditworthy actually repay the loan?''
Such observations naturally incorporate historic and on-going discrimination, as they usually do not ask: ``If a person who was labeled creditworthy did not repay the loan, was that because she can genuinely not handle money, or because she belongs to a disadvantaged group?'' To be sure, this question is crucial for the societal mitigation of discrimination in the long term. However, with the data we have at our disposition, we usually cannot answer it exhaustively. Hence, \methodname must generally content itself with correcting unequal calibration or disparate error rates, defined by a divergence from ground truth, that disproportionately affect a certain group of individuals.
As such disparities often correlate with the membership of a disadvantaged group in society, it may also correct discrimination caused by skewed mappings between construct and observable space to a certain extend (as we have seen in our experimental results). 
However, this is only a side-effect and depends on the nature of discrimination that is present in the data. If ground truth is close to reality, i.e., to construct space, then we can cure biased mappings between construct and observable space; otherwise, not. 
For the same reasons, \methodname addresses \emph{technical} and \emph{emergent}, but not \emph{pre-existing} bias, as defined by~\citet{friedman1996bias}.

We therefore advise \methodname to be used in settings where ground truth observations are sufficiently reliable (though this is a necessary condition for any data-driven decision making), and \emph{it can be assumed that groups can be meaningfully compared by the respective measures of qualification or quality}.
As an example for a non-comparable qualification measure across genders, consider the h-index \cite{hirsch2005index}, which suffers from strong gender bias.
Research shows that women, for the same amount and quality of research, receive significantly lower h-index values because it takes them longer to publish, and their work is less cited than men's~\cite{caplar2017quantitative}.
In cases where groups cannot be meaningfully compared, a method that implements statistical parity or other affirmative action policies independent of ground truth (such as FA*IR by \citet{zehlike2017fair, zehlike2022fair}) is an appropriate choice (see also~\cite{wachter2021bias}).

\subsection{Limitations of \methodname}

\paragraph{Continuous Mathematical Framework } 
As stated, we work with a \emph{continuous} scale of possible scores (in the real interval $[0,1]$) and accordingly with continuous probability distributions.
%For instance, $\nu_t([0.1,0.2])$ is the probability that a randomly chosen individual from group $t$ has score between $0.1$ and $0.2$ (where the probability is determined from historical data as explained in the preceding remark).
%
In practice, of course, there will be only a discrete set of possible scores, and a finite set of individuals under scrutiny.
The reason we work in a continuous setting is of a mathematical nature, in that it allows us to use certain tools from the theory of optimal transport, which in a discrete setting would complicate the theory considerably.
It is common and reasonable practice to approximate discrete models by continuous ones, as long as the number of individuals considered is large and their scores are sufficiently spread out over a finite scale.
Thus, our framework would not be suitable for a situation where there are only very few possible score values, e.g., only three or four.
Further discussion in this direction, for a related problem, can be found in~\citet{zehlike2020matching}.

\paragraph{Discontinuities in score distribution }
In criterion B) (we use this case as an example to explain the problem, but the same applies to criterion C), we only balance the scores for the negative class and ignore the true positives. This means that the scores of the true positives remain unchanged by \methodname. This might become a problem for rare scores of true negative individuals for whom there was no example in the data when calculating the optimal transport (OT) matrix. Remember that our algorithm calculates an optimal transport matrix to translate raw scores into fair scores for \emph{previously unseen} individuals. This means that for new data points, we do not know the ground truth label. Instead, \methodname consults the OT matrix that corresponds to a certain group and setting of $\theta$ and checks if it finds a raw score entry in that matrix.
In such a case, the following scenario might arise when trying to optimize for criterion B) only:
Suppose a true negative person's score lies at the edge of the overall true negative score distribution, i.e., only few others (in the world) have the same score while being true negative. In addition, critically, no true negative person had this score when the OT matrix was computed. In this case, the algorithm does not have an entry for translation in the OT matrix. As a result, this individual would be considered true positive, and their score would be left untouched.

\subsection{Critique of using COMPAS as an experimental scenario}
\label{sec:discussion:compas and missing data}
\citet{bao2021compaslicated} raise the problem that Risk Assessment Instrument (RAI) data sets such as COMPAS are used for benchmarking fairness methods in a purely technical fashion without recognizing the domain- and data-specific concerns of Criminal Justice (CJ). As a result, technical fairness method contributions could be interpreted as substantive interventions in the CJ domain. However, the authors rightly argue, the latter presupposes careful considerations of the domain specifics that may invalidate such results. They raise several serious concerns about COMPAS: outcome label bias due to construct invalidity; measurement bias that cannot be calibrated due to lack of (reliable) ground truth; protected attribute measurement bias; racial bias in covariates; and distribution issues due to label selection bias.

We recognize and acknowledge the concerns raised by Bao et al. However, since our contribution is, inter alia, an improvement on a previously published method presented on the COMPAS data set, we deem it important to enable a comparison of our results with the original. Thus, we publish the results from our experiments on the COMPAS data set only as a technical comparison with the results in \citet{kleinberg2016inherent}, and under the caveat that reliable ground truth is not available for the COMPAS data set. 
Hence, the results on applying FAIM to COMPAS are not intended as a contribution to real-world outcomes in the criminal justice domain. Rather, we discuss how \methodname could be used in a pretrial setting \emph{if} the model and data were considered adequate.

\subsection{Data Missingness}
Data missingness is a validity concern for the study of any fairness criterion, as noted by \citet{goel2021importance}, and is not unique for FAIM. As is the case for many fairness criteria, FAIM assumes the availability of ground truth data that is representative of the underlying distribution, and the absence of systematic censorship \cite{kallus2018residual} in the data generation process. If systematic censorship occurs due to biased decision making, then this prevents us from sampling from the ground truth distribution needed for an unbiased fairness criterion, thus making evaluation problematic.

Concerning the three data sets used in this paper: as mentioned, we use the COMPAS data set to allow a direct comparison with the metric proposed by \citet{kleinberg2016inherent}, and we thus do not want to modify the data set, for example through recovery or imputation. As shown by \citet{goel2021importance}, it is possible to recover parts of the distribution on certain assumptions about the causal structure of the decision-making process, but making such assumptions would not allow comparison with the results of \citet{kleinberg2016inherent} or allow us to make any general conclusions about applying FAIM to COMPAS. %Also, since this is not a contribution in the CJ domain we do not wish to make assumptions about the causal structure of the decision-making process that generated the COMPAS data set.

The e-commerce data set is subject to systematic censorship in the sense that articles classified as "low visibility" would receive less exposure and thus do not have opportunities to gain more exposure. Modifying the exploration-exploitation strategy for the learning to rank model would allow us to sample from the missing data to a larger degree, but performing such experiments would be prohibitively costly.

Finally, to further investigate the effect of data missingness on FAIM, we could introduce synthetic missingness to the synthetic data set, but would have to do so under the assumption of different causal structures, which are in an undefined and infinite search space. Thus, it is unclear what the relevant causal structures would be that would allow us to make meaningful conclusions.

\section{Conclusion}
As AI systems are deployed in an increasing number of societally relevant scenarios, considerations of fairness, non-discrimination and performance have assumed a central role in both in the policy debate surrounding AI and in AI research. However, implementations and operationalizations of these concepts are needed to ensure that AI systems can be deployed in ways that comply with, rather than violate, the law. In this paper, we have presented \methodname, a re-scoring algorithm that interpolates between three important criteria from the literature on fairness in machine learning, which also have immediate relevance for legal compliance questions: calibration, balance for the negative class, and balance for the positive class. Previous research has shown that they are mutually exclusive under reasonable factual assumptions. However, in many practical and legal scenarios, it is unsatisfactory -- and potentilly illegal -- to fully prioritize one criterion while neglecting the other two.

Our paper makes two main contributions. First, we move beyond the impossibility theorems that pit various competing notions of fairness against one another in a seemingly irreconcilable manner. \methodname allows to find an optimal compromise between three highly popular and important fairness criteria through three per-group parameters $\theta^A, \theta^B$, and $\theta^C$ using optimal transport theory. The parameter settings can be flexibly adapted to various scenarios. In this way, decision makers can balance competing interests and rights of protected groups; and they may adapt one model to different legal requirements.
We presented mathematical theory together with a pseudocode implementation to help readers with different backgrounds to follow our approach.
The model was tested in extensive experiments on real-world data sets from the criminal justice and the e-commerce domain.
Our experiments have shown that \methodname is effective under all scenarios, but that an inherently flawed model or data set should not be ``fixed'' via fairness constraints.
As such, we want to stress that our model is not a means of ethical reasoning, i.e., it can not determine whether a candidate should or should not receive bail or credit, or whether a product should be included in a top-$k$ ranking.
Instead, it ensures an optimal compromise for mutually exclusive fairness constraints according to the desires of a decision maker, expressed through the choices of $\theta^A, \theta^B$, and $\theta^C$.

Our second main contribution is to map these technical interpolation capacities to legal provisions requiring nuanced trade-offs between fairness, performance and other criteria relating to the protection of fundamental rights. Operationalizations of the often vague legal terms are crucial for compliance and to deploy AI systems in ways that are both safe and legal. We offer novel interpretations of the recently enacted EU AI Act and the Digital Markets Act against this technical background, and show how \methodname may help navigate these emerging, new frontiers at the intersection of law and computer science. \textcolor{maincolor}{Guidance is offered on how the trade-off between different fairness metrics may be conducted in various high-stakes settings, including hiring, college admissions, and healthcare.}

In doing so, we have also stressed that any choice of $\theta$ values must, of course, remain within the confines of the law. We discuss how \methodname may be harnessed to comply with competing legal non-discrimination obligations in three concrete settings corresponding to our experiments: credit scoring, bail decisions, and emerging notions of fairness in e-commerce rankings. The discussion shows that our model can handle both traditional legal fields and novel obligations under the recently enacted AI Act and the DMA. To our knowledge, our paper is the first to explore these new instruments from both a technical and a legal perspective.

\section{Acknowledgments}
We want to thank G\'eraud Le Falher and Pak-Hang Wong for their support and insightful suggestions to improve the paper.
\textcolor{maincolor}{We would like to acknowledge the authors of the Python Optimal Transport library~\cite{flamary2021pot} used extensively in our implementation of \methodname.}
The authors do not have any funding to disclose other than their affiliation.

This research did not receive any specific grant from funding agencies in the public, commercial, or not-for-profit sectors.

\bibliographystyle{plainnat}
%\bibliography{onsager}
\bibliography{00-Kleinberg_interpol}

\end{document}